\documentclass[a4paper,singlecoulomb,12pt]{article}
\usepackage{epsfig,amsmath,amssymb}
\usepackage[colorlinks=true,linktocpage=true,linkcolor=blue,citecolor=blue]{hyperref}
\usepackage{cite}
\usepackage{appendix}
\usepackage{graphicx}
\usepackage{dcolumn}
\usepackage{bm}
\usepackage{subfig}
\usepackage{float}
\usepackage{tabularx}
\usepackage{hhline}
\usepackage{color}
\usepackage{latexsym}
\newcommand{\ba}{\begin{eqnarray}}
\newcommand{\ea}{\end{eqnarray}}

\usepackage{caption}

\begin{document}
\title {\large Cumulative effects of collision integral,
strong magnetic field, and quasiparticle description on charge and 
heat transport in thermal QCD medium}

\author{Salman Ahamad Khan\footnote{skhan@ph.iitr.ac.in} and  Binoy Krishna
Patra\footnote{binoy@ph.iitr.ac.in} \vspace{0.3in} \\
Department of Physics, \\
Indian Institute of Technology Roorkee, Roorkee 247667, India}
\date{}
\maketitle
\vskip 0.01in

\begin{flushright}
{\normalsize
}
\end{flushright}

\begin{abstract} 

Our first aim is to explore the effect of the collision integral 
with the insurance of instantaneous conservation of particle number
on the charge and heat transport in a thermal QCD medium.
The second aim is to see how the dimensional reduction due to
strong magnetic field modulates the transport through the
entangled effects such as the collision-time and occupation
probability etc. in the collision integral. 
Final aim is to check how the quasiparticle description of partons,
through the dispersion relation of thermal QCD in strong magnetic field, 
alters the aforesaid conclusions. 
We observe that the modified collision term expedites both transport,
which is manifested by the larger magnitudes of electrical ($\sigma_{\rm 
el}$) and
thermal ($\kappa$) conductivities, in 
comparison to the relaxation collision term. As a corollary, the
Lorenz number is dominated by the later and 
the Knudsen number is by the former.  However, the strong magnetic field
not only flips the dominance of collision term in the heat transport, it
also causes drastic enhancement of both $\sigma_{\rm el}$ and $\kappa$ 
and reduction in the specific heat. As a result, the equilibration
factor, the Knudsen number
becomes much larger than one, which defies physical interpretation.
Finally, the quasiparticle description of partons in the
absence of strong magnetic field impedes the transport
of charge and heat, resulting in tiny decrease of the conductivities.
However, the strong magnetic field does noticeable observations: 
the conductivities now gets reduced to the physically plausible values, 
the temperature dependence of $\sigma_{\rm el}$ gets reversed, {\em i.e.} 
it now decreases with temperature, effect of collision integral
gets smeared in $\kappa$ etc. The Knudsen number thus
becomes much smaller than one, implying that the system be remained
in equilibrium. These findings attribute to the fact that the collective
oscillation in the dispersion relation of thermal QCD 
in strong magnetic field sets in much larger scale, manifested 
by the large in-medium flavour masses.
\end{abstract}

\section{Introduction}
Quark Gluon Plasma (QGP), a deconfined state of
quarks and gluons is formed in heavy ion 
collision experiments at Relativistic Heavy Ion collider (RHIC) 
and Large Hadron Collider (LHC). It is believed that our present universe 
was also in QGP form around one microsecond after the big bang. There exists 
such hot and 
dense matter in the core of the neutron stars also. In the
ongoing heavy ion collision experiments 
at RHIC and LHC, a very strong magnetic field, 
perpendicular to the reaction plane, could be produced in the 
very early stages of the collisions in noncentral 
events\cite{Fukushima,Muller:PRD89'2013} 
{\em viz.} order of $m_\pi^2$ at RHIC
\cite{Kharzeev:NPA803'2008} and $15m_\pi^2$ at 
LHC \cite{Skokov:IJMPA24'2009}.
Some very naive 
estimates predict that the decay of the magnetic 
field in the non-conducting medium is very fast but due  to the finite 
value of the electrical conductivity of the medium, the decay time 
for the magnetic field gets elongated, which may then cause to affect
the physical quantities associated with QGP. In the recent
years, the effects of the background magnetic field on the 
various properties of the QGP have been investigated by the various 
research groups, {\em such as} the chiral magnetic 
effect\cite{Fukushima:PRD78'2008,Kharzeev:NPA803'2008},
magnetic and inverse magnetic catalysis\cite{Gusynin:PRL73'1994,Leung:PRD55'1997,
Gusynin:PRD56'1997,Shovkovy:LNP'871'2013}, 
axial magnetic effect\cite{Braguta:PRD89'2014,
Chernodub:PRB'2014},
chiral vortical effect in rotating QGP
\cite{Kharzeev:PRL106'2011, Kharzeev:PPNP88'2016},
the conformal anomaly
 and production of soft photons\cite{Fayazbakhsh:PRD88'2013,
 Basar:PRL109'2012},
 apart from that the dilepton production rate\cite
{Tuchin:PRC88'2013,Bandyopadhyay:PRD94'2016,Sadooghi:ANP376'2017}, 
 dispersion relation in the magnetized thermal 
QED\cite{Sadooghi:PRD92'2015}, refractive indices 
and decay constants\cite{Fayazbakhsh:PRD86'2012,
Fayazbakhsh:PRD88'2013} and various thermodynamical 
properties  \cite{Rath:JHEP1712'2017,Karmakar:PRD99'2019}.

In the strong magnetic field ($|q_iB|>>T^2 $ and $|q_iB|>>m^2_i$,
where $q_i (m_i)$ is the electric charge (mass) of the $i$-th flavour),
the dynamics of charged particles are constrained only in the
lowest Landau levels, so the dispersion relation ($E=\sqrt{p_3^2+m_i^2}$,
involves only the momentum ($p_3$) along the direction of magnetic field.
Some recent observations \cite{Tuchin:AHEP2013'2013,Gursoy:PRC89'2014} indicates
the similarity of the time scales between the formation of the 
locally equilibrated thermal QCD medium and  the production of strong
magnetic field, due to faster thermalization. Thus, the
transport properties of the medium may be 
affected due to presence of the strong magnetic field. Electrical 
conductivity $(\sigma_{el})$, which is responsible for the generation of electric
current due to Lenz's law plays a vital role in the study
of the chiral magnetic effect \cite{Fukushima:PRD78'2008}. 
Apart from that, $\sigma_{el}$ plays very crucial role as an input 
parameter in the phenomenological studies at RHIC 
and LHC, {\em such as} the emission rate of soft photons \cite{Kaputa:CUP'2006}.
The effect of the magnetic field on the electrical conductivity
has been previously studied using different approaches, {\em such as} 
the dilute instanton-liquid model \cite{Nam:PRD86'2012}, the nonlinear
electromagnetic currents \cite{Kharzeev:PPNP75'2014,
Satow:PRD90'2014}, the diagrammatic method using real time formalism
\cite{Hattori:PRD94'2016}, quenched SU(2) lattice gauge theory
\cite{Buividovich:PRL105'2010}, axial Hall current \cite{Pu:PRD91'2015}, the 
effective fugacity approach \cite{Kurian:PRD96'2017} etc. 

Another important transport coefficient is the thermal conductivity ($\kappa$),
which is the measure of a medium to conduct the heat through it.
The thermal dissipation in the medium depends on the temperature
gradient associated with the different layers  of the fluid.
The studies on $\kappa$ of the hot QCD matter in a strong magnetic field 
have recently been done in ~\cite{Kurian:PRC79'2019,
Rath:PRD100'2019}. In fact, $\kappa$ plays a crucial role 
in terms of Knudsen number to check that the system is in 
local equilibrium. The Knudsen number is the ratio
of the mean free path $(\lambda)$ to the characteristic length of the
system ($L$), where $\lambda$ is related to $\kappa$ by
$\lambda= 3\kappa/({\mathrm v C_V})$ (where ${\rm v}$ is the 
relative velocity of quark and ${\mathrm C_V}$ is the specific heat at 
the constant volume). The system is said to be in 
local hydrodynamic equilibrium if the mean free path is smaller 
than the characteristic length of the medium. The $\sigma_{{\rm el}}$ 
and $\kappa$ are not independent rather their ratio $\kappa/\sigma_{el}$ 
is equal to the 
product of Lorentz number $L_R$ and the temperature, this relation 
is  commonly known as the Wiedemann-Franz Law. Metals being good 
conductor of heat and electricity follow Wiedemann-Franz law 
perfectly. However, violation of the this law has been recorded
in many systems such as, the
two-flavor quark matter in the Nambu-Jona-lasinio model
\cite{Harutyunyan:PRD95'2017}, the strongly interacting QGP medium \cite{Mitra:PRD96'2017}, thermally 
populated electron-hole plasma in graphene \cite{Crossno:Scince351'2016} 
the unitary Fermi gas \cite{Husmann:PNAS115'2018,Han:PRA100'2019} 
and the hot hadronic matter \cite{Rath:EPJA55'2019}.

In the previous studies \cite{Rath:PRD100'2019,Rath:EPJC80'2020}, 
the transport coefficients have been calculated from the Boltzmann
equation and the complexities of the collision term was avoided by
a mean-free-path treatment. In this treatment, the collision integral
is simplified by a relaxation term, implying that the collisions
tend to relax the distribution function to an equilibrium value.
The relaxation model then describes the destruction of phase of
an ordered motion on collision and leads to a damping frequency of
order $1/\tau$ in the amplitude, where $\tau$ is some suitable average
collision time. This type of model has a flaw that charge is
not conserved instantaneously but only on the average over a cycle.
It is, however, easy to remedy this at least in the case of
constant collision time by modifying the collision term due to
Bhatnagar-Gross-Krook (BGK)~\cite{Bhatnagar:PRD94'1954}. 
The BGK collision term differs 
physically from the relaxation type in the following manner:
Each collision term in a Boltzmann equation consists of two parts, where the
first one represents particles removed or absorbed 
from a definite momentum range by collisions and the second one 
represents the particles emitted into that range as a result of collisions. 
The absorption term is essentially the same as that in the 
relaxation term, {\em i.e.} particles in
a momentum range $d\vec{p}$ about the momentum $\vec{p}$ are absorbed 
at a rate proportional to perturbed distribution $f(\vec{x},\vec{p},t)$.
The emission term is the real source of difficulty,
for which BGK prescribed that particles emitted at a rate proportional
to the product of perturbed particle density, $n (\vec{x},t)$ and 
the equilibrium distribution function~\cite{Bhatnagar:PRD94'1954}. 
The effect of the BGK  collision term which ensures the 
conservation of the particle number instantaneously has been studied on 
the plasma instabilities in ~\cite{Schenke:PRD73'2006}.
 

Thus, the aim of the present work is to extend/modify of the aforesaid
recent works\cite{Rath:PRD100'2019} on the transport of charge and 
heat in a hot quark matter 
in three fold respects: (i) By modifying the collision terms as 
mentioned above, the solution of Boltzmann equation for the 
infinitesimal disturbance gets altered, which, in turn, affects
the transport coefficients directly.
(ii) The effect of a strong magnetic field, a possibility 
in the peripheral events of ultrarelativistic heavy-ion collisions, 
is explored, due to the reduction of phase space and the enhancement
of the collision time etc.
(iii) Finally the role of interactions is explored
in the quasiparticle description of partons,
where the vacuum masses of partons are replaced by the masses 
generated in the medium. These masses are obtained from the
pole of full propagator, calculated by the 
perturbative thermal QCD in the background
of strong magnetic field.
Thus, we have employed a kinetic theory approach with the
BGK Collision term in Boltzmann equation to compute 
the electrical ($\sigma_{el}$) and thermal ($\kappa$) conductivities 
and the derived coefficients (Lorenz and Knudsen number) from them.

We have found that the modified collision integral enhances 
the magnitudes of both conductivities, especially more
to the electrical conductivity, compared to the counter-parts
with the collision term of relaxation type. 
As a consequence, the ratio, $\kappa/\sigma_{\rm el}$ and the 
Lorenz number, $L_R$ $ (=\kappa/\sigma_{el}T)$ gets decreased
whereas the the equilibration factor, Knudsen number gets
increased. In the presence of strong magnetic field, interesting
thing happens in the transport. Although there is an overall enhancement
of both conductivities but $\kappa$ becomes smaller
than in relaxation collision integral. This
could be attributed to the constrained motion of quarks in
strong magnetic field. As a corollary, the Knudsen number
becomes much larger than one, due to the enhancement of $\kappa$
and the reduction of specific heat, thus necessitates the quasiparticle
description (QPD) of partons for the transport phenomena, at least,
in strong $B$. In fact, the QPD of partons (mainly for quarks) in the absence
of magnetic field makes the transport coefficients a little bit smaller
but in the presence of strong magnetic field, QPD enriches the transport
phenomena  interesting: $\sigma_{el}$ now decreases with temperature
and $\kappa$ becomes insensitive to the collision integral, except the 
temperature is very high. As a result, the Knudsen number does
not depend on the type of collision integral and 
becomes much smaller than one, which ensures that the system is
still in local equilibrium even in the presence of strong $B$.
This could be envisaged by the fact that the collective oscillation
of a thermal QCD medium
in strong $B$ sets in much bigger scale than in the absence
of magnetic field. 

The paper has been organized as follows:  In section 2, we have 
investigated the effect of the collision term via the collision integral 
and the effect of (strong) magnetic field on the charge and 
heat transport, separately, in a thermal medium of noninteracting 
quarks and gluons. For that purpose, we have employed
the kinetic theory approach through 
the relativistic Boltzmann transport equation.
In particular, subsections 2.1 and 2.2 deal with
the charge and heat transport via the computation of 
electrical and thermal conductivities, respectively. The subsection
2.3 explores the transport coefficients related the competition
between thermal and charge transport through the ratio, $\kappa/\sigma_{el}$ 
(and the Lorenz number, $L_R$ in Wiedemann-Franz Law) and the validity 
of local equilibration through the Knudsen number. 
Thus, having explored the sole effect of collision terms on the charge and heat
transport, we explore
the sole effect of background strong magnetic field on the aforesaid
transport coefficients in subsections 2.4 and 2.5, respectively.
We have 
found that the magnitude of conductivities for noninteracting partons
in strong magnetic field defy physical interpretation,
contrary to the spirit of the linearization of the collision integral 
on the basis of near-equilibrium assumption.
This motivates us to compute the same in section 3
with the interactions present among the
constituents of the medium in the guise of 
quasiparticle description of partons. So we have first 
revisited the medium generated thermal masses for 
the quark flavors in the presence of the strong magnetic field in 
subsection 3.1
and then investigate how the quasiparticle description in section
3.1 affects the abovementioned transport phenomena and found 
the plausible magnitudes of the conductivities and other
derived transport coefficients. Finally Section 4 concludes results
and discussions. 

\section{Charge and Heat Transport in a thermal medium of noninteracting 
quarks and gluons} 
The Boltzmann transport equation for a single particle distribution function is 
given by
\begin{eqnarray}
\frac{\partial f}{\partial t} +\frac{\mathbf p}{m} \cdot \nabla f +
{\mathbf F} \cdot \frac{\partial f}{\partial \mathbf p}= {\left(
\frac{\partial f}{\partial t} \right)}_{\rm coll},
\label{bte}
\end{eqnarray}
where ${\mathbf F}$ is the force field acting on the particles
in the medium and the term on the right hand side is added to describe the 
effect of collisions between particles. If the collision term is zero 
then the particles do not collide, where individual collisions get
replaced with long-range aggregated Coulomb interactions, referred 
as the collisionless Boltzmann equation or Vlasov equation.

The solution of the Boltzmann equation is, in general, a matter of 
considerable difficulty even in the cases corresponding to the
physically simplest situations. The main difficulty in handling the full
Boltzmann equation arises from the complicated nature of the collision terms,
consisting of absorption and emission terms. The absorption
causes the removal of particles from a definite momentum range by 
collisions and then the particles are emitted into that range as a result 
of collisions. The absorption term is substantially the same as that in the 
Boltzmann equation, {\em i.e.} particles in
a momentum range $d\vec{p}$ about the momentum $\vec{p}$ are absorbed 
at a rate proportional to perturbed distribution $f(\vec{x},\vec{p},t)$.
The emission term is the real source of difficulty for which we will now
discuss some simple kinetic models, which permit of exact mathematical
treatment including the solution of definite boundary value problems.

In many kinetic problems, it is convenient to avoid the complexities
of the Boltzmann equation by using a mean free-path treatment, where one
replaces the collision integral by a relaxation term of the form
\ba
{\left( \frac{\partial f}{\partial t} \right)}_{\rm coll} 
= - \frac{1}{\tau(p)} \left(f(\vec{x},\vec{p},t)-f_{\rm eq} (|\vec{p}|)
\right),
\label{RT}
\ea
where $\tau$ is the momentum-dependent collision time, which
implies that the collisions tend to relax the distribution function
to an equilibrium value $f_{\rm eq} (|\vec{p}|)$, which is a function
of momentum only. We illustrate the collision models by 
referring to oscillatory problems, where a characteristic time enters 
in a natural
way. The above collision term \eqref{RT} of relaxation type then describes 
the destruction of phase of
an ordered motion on collision and leads to a damping frequency of
order $1/\tau$ in the amplitude, where $\tau$ is some suitable average
collision time. This type of model has the defect that the charge is
not conserved instantaneously but only on the average over a cycle.
It was first remedied by Bhatnagar-Gross-Krook (BGK)~\cite{Bhatnagar:PRD94'1954}
, where the particles 
in a range $d^3p$ about momentum $\vec{p}$ are absorbed at 
a rate proportional to the number $f(\vec{p}, \vec{x},t)$ at $(\vec{x},t)$, 
and are re-emitted at a rate proportional to the perturbed density, 
$n(\vec{x},t)$ (=$\int f(\vec{x},\vec{p},t) d^3p$). The BGK
collision term is given by \cite{Schenke:PRD73'2006,Bhatnagar:PRD94'1954}

\begin{eqnarray}
{ \left( \frac{\partial f}{\partial t}\right)}_{\rm coll} = -\frac{1}{\tau} 
\left(f(\vec{x},\vec{p},t)- \frac{n(\vec{x},t)}
{n_{\mathrm eq}} f_{\rm eq} (|\vec{p}|)\right),
\label{rbte}
\end{eqnarray}
which upon  integrating over momenta vanishes {\em i.e.} 
it conserves the particle number instantaneously.
 Its effect was extensively discussed 
on QCD plasma instabilities~\cite{Schenke:PRD73'2006}.

Till now we have discussed the nonrelativistic version of the Boltzman 
equation, however, the transport phenomena for a medium consisting of 
quarks and gluons could be better understood by the relativistic 
generalization of Boltzmann equation, which is oftenly
expressed in a covariant form as
\begin{eqnarray}
p^{\mu}\frac{\partial f}{\partial x^{\mu}}+q~F^{\rho \sigma}
p_{\sigma}\frac{\partial f}{\partial p^{\rho}}= 
-\frac{p^\mu u_\mu}{\tau}
\left(f(\vec{x},\vec{p},t)- \frac{n(\vec{x},t)}
{n_{\mathrm eq}} f_{\rm eq} (|\vec{p}|)\right),
\label{rbte}
\end{eqnarray}
where $F^{\mu \nu}$ is the external electromagnetic force.
We will now see in forthcoming sections how the abovementioned collision 
integral affects the solution of the relativistic Boltzman equation, 
which, in turn, alters the transport of electricity and heat in 
terms of their respective
transport coefficients, {\em such as} the electrical and thermal 
conductivities and
the derived coefficients from them, namely Lorenz and Knudsen number. 
Furthermore we also explore how a strong magnetic field could 
modulate the effect of modified collision term on the aforesaid transport
processes. 

\subsection{Electrical conductivity in the absence of magnetic field}
The linear response of a medium consisting of mobile charge carriers to an 
infinitesimal electric field (${\bf E}$) deciphers the electrical properties 
of the medium. The electric field induces a current ($\vec{J}$) in
the medium, which is proportional linearly to the (infinitesimal)
applied ${\bf E}$ and the proportionality constant, known as the electrical
conductivity ($\sigma_{\rm el}$), determines the electrical properties
of the given medium. For a thermal QCD medium consisting of quarks, 
antiquarks and gluons of different flavours ($i$), the four-component 
induced current, contributed by quarks and antiquarks only, becomes 
\ba
J_{\mu}= \sum_i q_ig_i \int 
\frac{d^3p}{(2\pi)^3}~\frac{p_{\mu}}{\omega_i} 
~(\delta f_i(x,p)+\delta \bar{f}_i(x,p))
\label{current_temp}
\ea
where, $q_i,$ $g_i$ and $\delta f_i(x,p)$ are the charge, 
degeneracy factor and the infinitesimal deviation from the equilibrium 
distribution function for ${\rm i}$-th quark, respectively. Similarly
$\delta \bar{f}_i (x,p)$ denotes for ${\rm i}$-th anti-quarks, which, 
for vanishingly small quark chemical potential ($\mu \approx 0$), 
is the same as for quarks, $\delta f_i(x,p)$. Therefore, the
induced current can be calculated provided the infinitesimal
deviation, $\delta f_i$ is known. In kinetic theory 
approach, the $\delta f_i$ is obtained from the
solution of relativistic Boltzman equation, after linearizing the
collision integral with respect to a (infinitesimal) 
perturbation to a medium, which was initially in equilibrium.

In order to see the responce of the electric field, we take 
only $\rho =i$ and $\sigma=0$ and {\em vise versa} components of the 
electromagnetic field strength tensor, {\em i.e.}  $F^{0i}=-{\bf E}$ 
and $F^{i0}={\bf E}$ in the relativistic Boltzman equation
(RTBE) in \eqref{rbte}. Hence, the RTBE \eqref{rbte} gets reduced for the
$i$-th species in a multicomponent medium
\ba
\label{rbte_sigmatemp}
q_i {\bf E.p} \frac{\partial f_i}{\partial p^0}+
q_i p_0 {\bf E.}\frac{\partial f_i}{\partial {\bf p}}=C[f].
\label{BGK_sigma}
\ea
The modified collision integral $C[f]$ in \eqref{BGK_sigma} due to BGK
is generalized for the $i$-th species of a multicomponent system as
\ba 
\label{BGK_thermal}
C[f]=-p^{\mu}u_{\mu}\nu_i \left(f_i-n_{{}_i}n_{\mathrm eq,i}^{-1} f_{\mathrm eq,i}
\right), 
\ea
where $f_{\mathrm eq, i}$ is the equilibrium
distribution function of $i$-th flavour:
\ba
f_{\mathrm eq, i} = \frac{1}{e^{\beta u^\alpha {p_{\rm i}}_\alpha}+1},
\label{dis_quark}
\ea
where $p_{\rm i}^\alpha$ is $(\omega_i, \vec{p})$
and $u_\alpha$ is the
fluid four-velocity, which, in the local rest frame, is
$u_{\mu}=(1,0,0,0)$. The collision frequency, $\nu_i$ is the inverse of 
the relaxation-time of the medium, $\tau_i$. The relaxation time
can also be calculated from the Boltzmann equation, where
the gluon-gluon collision mainly plays the dominant role
in the collision integral, to bring the perturbed
system back to the equilibrium. The expression for $\tau$ 
is given by \cite{Hosoya:NPB250'1985}
\ba
\tau_i(T) =\frac{1}{5.1T \alpha_s^2 \log(\frac{1}{\alpha_s})
\label{tau_B0} 
[1+0.12(2N_i+1)]},  
\ea 
where $\alpha_s$ is the running coupling constant, which runs with the 
temperature as
 \begin{eqnarray}
\alpha_s (T)=\frac{6\pi}{(33-2N_f) \ln(\frac{Q}{\Lambda_{QCD}})},
\label{coupling_T}
\end{eqnarray} 
where $Q$ is set at $2\pi T$. \\

The symbol, $n_i$ in the above collision integral \eqref{BGK_thermal} 
represents the perturbed density for $i$-th species
\ba
n_i &=& g_i \int \frac{d^3p}{(2\pi)^3} (f_{\mathrm eq,i}+\delta f_i),
\ea
and the equilibrium density for $i$-th flavour, having degeneracy
factor, $g_i$, is given by
\ba
n_{\mathrm eq,i} = g_i \int \frac{d^3p}{(2\pi)^3}f_{\mathrm eq,i}.
\ea

After linearizing the collision integral with respect to the infinitesimal
perturbation: $f_{\mathrm eq,i} \rightarrow f_{\mathrm eq,i}
+\delta f_i,~  \delta f_i \ll f_{\mathrm eq,i}$, the RTBE \eqref{BGK_sigma}
is recast into the form \footnote{Using the symbol for momentum
integration, $\int_p = \int d^3p/{(2 \pi)}^3$.} 
(Details are given in Appendix A)
\ba \label{appendix_A}
\delta f_i- g_in_{\mathrm eq,i}^{-1} f_{\mathrm eq,i} \int_{p}\delta f_i=
2q_i\beta \tau_i~ \frac{{\bf E \cdot p}}{\omega_i} f_{\mathrm eq,i}\left( 
1-f_{\mathrm eq,i} \right),  
\ea
wherein the following partial derivatives have been used:
\ba \label{partial11}
\frac{\partial f_{\mathrm eq,i}}{\partial p^0}&=&-\beta f_{\mathrm eq,i}
(1-f_{\mathrm eq,i}),\\
\label{partial22}
\frac{\partial f_{\mathrm eq,i}}{\partial {\bf p}}&=&-\frac{\beta{\bf p}}{\omega_i}
 f_{\mathrm eq,i}(1-f_{\mathrm eq,i}).
\ea 
Therefore, the solution for $\delta f_i$ is obtained (neglecting the
higher order, ${\cal O} { (\delta f_i)}^2$) as
\begin{eqnarray}
\delta f_i=\delta f_i^{(0)}+g_i n_{\mathrm eq,i}^{-1} f_{\rm eq,i} 
\int_{p'}\delta f_i^{(0)},
\label{eqoneseven}
\end{eqnarray}
where
\begin{eqnarray}
\delta f_i^{(0)}=\frac{2\beta q_i \tau_i}
{\omega_i} {\bf E.p} ~f_{\mathrm eq,i}(1-f_{\mathrm eq,i}).
\end{eqnarray}

Thus, the spatial component of the four-vector for the induced
current is finally obtained by plugging $\delta f_i$ into 
\eqref{current_temp}, 
\ba
J_k&=& 4\beta \sum_i q_i^2 g_i \tau_i \left[\int 
\frac{d^3p}{(2\pi)^3}~\frac{p_k^2}{\omega_i^2(p)} 
  f_{\mathrm eq,i}(p)(1-f_{\mathrm eq,i}(p))\right. \nonumber\\
&&\left. +g_i n_{\mathrm eq,i}^{-1} \int_p~
\frac{p_k}{\omega_i(p)} 
f_{\mathrm eq,i}(p)\int_{p'}
\frac{ p'_k}{\omega_i(p')} f_{\mathrm eq,i}(p') (1 - 
f_{\mathrm eq,i}(p'))\right]E_k.
\ea
Hence, the coefficient of ${\bf E}$ in the above induced current,
${\bf J}$ thus yields the electrical conductivity in the modified
collision term which can be decomposed in terms of the contribution due 
to the collision term of the relaxation type \eqref{RT} and a correction term as
\ba \label{sigma_bgk}
\sigma_{\rm el}=\sigma_{\rm el}^{\rm RT}+\sigma_{\rm el}^{\rm Corr}, 
\ea


where
\ba
\sigma_{\rm el}^{\rm RT} &=& 4\beta \sum_i q_i^2 g_i \tau_i 
\int_p ~\frac{p^2}{3\omega_i^2(p)} 
  f_{\mathrm eq,i}(p)(1-f_{\mathrm eq,i}(p)),\\
\sigma_{\rm el}^{\rm Corr} &=& 4\beta \sum_i q_i^2 g^2_i \tau_i n_{\mathrm eq,i}^{-1} 
\int_p~\frac{p}{\omega_i} 
f_{\mathrm eq,i}(p)\int_{p^\prime} \frac{d^3p'}{(2\pi)^3}
\frac{ p'}{\omega_i(p')} f_{\mathrm eq,i}(p') (1 - 
f_{\mathrm eq,i}(p')),
\label{sigmaRT}
\ea
where $\sigma_{\rm el}^{\rm Corr}$ is found to be positive,
implying that the modified BGK collision term always enhances the charge
transport.

\subsection{Thermal conductivity in the absence of magnetic field}
In this section, we will calculate the thermal conductivity from
the surplus of the energy diffusion over the enthalpy diffusion,
known as the heat flow. In four-vector notation, the heat-flow
is defined as 
\begin{eqnarray}
Q_{\mu}=\Delta_{\mu\alpha}T^{\alpha\beta}u_{\beta}
-h\Delta_{\mu\alpha}N^{\alpha},
\label{Qfourvector}
\end{eqnarray}
where the projection tensor, $\Delta_{\mu \alpha}$ is given by 
\ba
\Delta_{\mu\alpha}=g_{\mu\alpha}-u_{\mu}u_{\alpha}, 
\ea 
and the enthalpy per particle, $h$ is 
\ba
h=(\varepsilon+P)/n,
\ea
where $\varepsilon$, $P$, and $n$ are the energy, pressure, and particle 
number densities, respectively. 
$N^{\alpha}$ and $T^{\alpha\beta}$  
are the particle flow number and the energy-momentum tensor 
(also known as the first and second moment of the 
distribution function, respectively), respectively, and are
defined in the kinetic theory for a multi-component system as
\begin{eqnarray}
N^{\alpha}&=&\sum_i 2g_i \int \frac{d^3p}{(2\pi)^3}
~\frac{p^{\alpha}}{\omega_i}~
f_{\mathrm eq,i},\\
T^{\alpha\beta}&=&\sum_i 2g_i \int \frac{d^3p}{(2\pi)^3}
\frac{p^{\alpha}p^{\beta}}{\omega_i}~f_{\mathrm eq,i},
\end{eqnarray}
which yield $n$, $\varepsilon$ and $P$ by the following contractions: 
\begin{eqnarray}\label{numdensity1}
n &=&N^{\alpha}u_{\alpha},\\
\label{energyden1}
\varepsilon &=& u_{\alpha}T^{\alpha\beta}u_{\beta},\\
\label{pressuredendity1}
P&=&-\frac{\Delta_{\alpha\beta}T^{\alpha\beta}}{3},
\end{eqnarray}
respectively.\\ 

In the rest frame of the fluid, the heat flow four vector 
is orthogonal to the fluid four-velocity
\ba
 Q^{\mu}u_{\mu}=0,
\ea
so, both temporal and spatial components are not independent rather
the heat four vector can be determined by its spatial 
component alone. Thus, the spatial component is read off from 
\eqref{Qfourvector} and can be expressed  in kinetic theory 
for a multi-component system as
\begin{eqnarray}
{\bf Q}=\sum_i 2g_i \int \frac{d^3p}{(2\pi)^3}~\frac{\bf p}
{\omega_i}~(\omega_i-h_i)~\delta f_i,
\label{heat_vec1}
\end{eqnarray}
where, the enthalpy per particle for the $i^{th}$ flavour is, 
\ba
h_i&=&\frac{(\epsilon_i+ P_i)}{n_{\mathrm eq,i}}
\ea
with 
\ba
\epsilon_i &=& g_i\int \frac{d^3p}{(2\pi)^3}\omega_i f_{\mathrm eq,i} \\
P_i &=&\frac{g_i}{3} \int \frac{d^3p}{(2\pi)^3} \frac{p^2}{\omega_i} 
f_{\mathrm eq,i}
\ea
respectively.

In order to understand the dissipative processes in 
a medium in kinetic theory approach, {\em namely} the thermal 
conduction and viscosity etc., 
one usually goes to the next approximation beyond the initial local
equilibrium distribution function: $f_i=f_{\mathrm eq,i}+ \delta f_i$
($\delta f_i$ $\ll f_{\mathrm eq,i}$) and $\delta f_i$ is 
thereafter obtained from the Boltzmann equation \eqref{rbte}, 
after linearizing the collision integral with respect to the 
deviation.  

So we start with rewriting the RTBE \eqref{rbte} in a suitable form through the 
chain rule of differentiation
\begin{eqnarray}
p^{\mu} \frac{\partial f_i}{\partial T} \frac{\partial T}{\partial x^{\mu}}
+p^{\mu} \frac{\partial f_i}{\partial p^0}
\frac{\partial (p^{\nu} u_{\nu})}{\partial x^{\mu}}
+q_i\left[F^{0j}{\bf p}
\frac{\partial f_i}{\partial p^0}+F^{j0}p_0\frac{\partial f_i}{\partial p^j}
\right]=C[f].
\label{rbte_tempheat}
\end{eqnarray}
Now, going beyond the initial local equilibrium distribution function,
we first compute the left hand side of RTBE \eqref{rbte_tempheat} as
\begin{eqnarray}
{\rm LHS}&=&(p^{0}\partial_{0}T+p^{j}\partial_{j}T)\frac{\partial 
f_{\mathrm eq,i}}{\partial T}
+(p^{\mu} u_{\nu}\partial_{\mu}p^{\nu}+p^{\mu} p^{\nu}\partial_{\mu}u_{\nu})
\frac{\partial f_{\mathrm eq,i}}{\partial p^0}+q_i\left[{\bf E.p}
\frac{\partial f_{\mathrm eq,i}}{\partial p^0}+{\bf E}p_0\frac{
\partial f_{\mathrm eq,i}}{\partial p^j}
\right],\nonumber \\
&=&\frac{p^0}{T}f_{\rm eq,i}(1-f_{\rm eq,i})\left[\frac{1}{T}(p^{0}
\partial_{0}T+p^{j}\partial_{j}T)
-\frac{1}{p^0}(p^{\mu} u_{\nu}\partial_{\mu}p^{\nu}
+p^{\mu} p^{\nu}\partial_{\mu}u_{\nu})-2q_i
\frac{{\bf E. p}}{p^0}\right],\nonumber\\
&=&\frac{p^0}{T}f_{\rm eq,i}(1-f_{\rm eq,i})\left[\frac{1}{T}(p^{0}
\partial_{0}T+p^{j}\partial_{j}T)
-\frac{1}{p^0}(p^{0} \partial_{0}p^{0}+p^{j}
 \partial_{j}p^{0})
 -\frac{1}{p^0}(p^{0} p^{\nu}\partial_{0}u_{\nu}+p^{j} p^{\nu}
\partial_{j}u_{\nu})\right.\nonumber\\
&&\left. -2q_i\frac{{\bf E.p}}{p^0}\right]\nonumber\\
&=&\frac{p^0}{T}f_{\rm eq,i}(1-f_{\rm eq,i})\left[\frac{1}{T}
(p^{0}\partial_{0}T+p^{j}\partial_{j}T)
+T\partial_0\left(\frac{\mu}{T}\right)+\frac{T}
{p^0}p^j\partial_j\left(\frac{\mu}{T}\right)
 -\frac{1}{p^0}(p^{0} p^{\nu}\partial_{0}
u_{\nu}+p^{j} p^{\nu}\partial_{j}u_{\nu})\right.\nonumber\\
&&\left. -\frac{2q_i{\bf E.p}}{p^0}\right].
\label{lhs_heat}
\end{eqnarray}

The energy-momentum conservation facilitates to calculate the partial 
derivatives appeared in the above equation as
\ba
\partial_j\left(\frac{\mu_i}{T}\right)&=&-\frac{h_i}{T^2}
\left(\partial_jT-\frac{T}{n_{\mathrm eq,i} h_i}\partial_jP\right)\\
\partial_0u_{\nu}&=&\frac{\nabla_{\nu}P}{n_{\mathrm eq,i}h_i}.
\ea

Thus, the RTBE \eqref{rbte_tempheat} is written
by linearizing the BGK collision term \eqref{BGK_thermal}
\begin{eqnarray}
\frac{p^0}{T}f_{\rm eq,i}(1-f_{\rm eq,i})\left[\frac{p^0}{T}\partial_{0}T
+\frac{(p^0-h_i)}
{p^0}\frac{p^j}{T}\left(\partial_jT-\frac{T}{n_{\mathrm eq, i} h_i}
\partial_j P\right)
+T\partial_0\left(\frac{\mu_i}{T}\right)
-\frac{p^jp^{\nu}}{p^0} \partial_{j}u_{\nu}\right.\nonumber\\
\left. -2q_i\frac{{\bf E.p}}{p^0}\right]
=-p^{\mu}u_{\mu}\nu_i\left(\delta f_i-
g_in_{\mathrm eq,i}^{-1} f_{\rm eq,i}\int_p \delta f_i\right),
\label{rbte_heat}
\end{eqnarray}
which is further solved to obtain $\delta f_i$ (neglecting
its higher orders)
\begin{eqnarray}
\delta f_i=\delta f_i^{(1)}+g_in_{\mathrm eq,i}^{-1} f_{\rm eq,i}
\int_{p'} \delta f_i^{(1)},
\end{eqnarray}
where 
\begin{eqnarray}
\delta f_i^{(1)}&=&-\frac{f_{\rm eq,i}(1-f_{\rm eq,i})\tau_i}{T}
\left[\frac{p^0}{T}\partial_{0}T+\frac{(p^0-h_i)}
{p^0}\frac{p^j}{T}\left(\partial_j T-\frac{T}{n_{\mathrm eq,i} h_i}
\partial_jP\right)
+T\partial_0\left(\frac{\mu_i}{T}\right)\right.\nonumber \\
&&\left. -\frac{p^jp^{\nu}}{p^0} \partial_{j}u_{\nu}
-2q_i\frac{{\bf E.p}}{p^0}\right].
\end{eqnarray}

Thus, the spatial part of the heat flow vector is obtained
by plugging $\delta f_i$ into the equation \eqref{heat_vec1}.
\begin{eqnarray}
{ Q_j} &=& \sum_i 2g_i\int \frac{{d^3p}}{(2\pi)^3}~\frac{{ p_j}}
{\omega_i(p)}~(\omega_i(p)-h_i)\left[\left\{\frac{f_{\mathrm eq,i}(p)
(1-f_{\mathrm eq,i}(p))\tau_i}{T}
\frac{(p^0-h_i)}{p^0}\frac{p_j}{T}\right. \right. \nonumber\\
&& \left. \left. +g_in_{\mathrm eq,i}^{-1} f_{\mathrm eq,i}(p)\int_{p'}
\frac{f_{\mathrm eq,i}(p')(1-f_{\mathrm eq,i}(p'))\tau_i}{T}\frac{(p^0-h_i)}
{p^0}\frac{p_j}{T}\right\} \left(\partial_ j T-\frac{T}{nh_i}
\partial_j P\right)\right.\nonumber \\ 
&& \left. +\left(\frac{f_{\mathrm eq,i}(p)(1-f_{\mathrm eq,i}(p))\tau_i}{T}
+ g_in_{\mathrm eq,i}^{-1} f_{\mathrm eq,i}(p)\int_{p'}
\frac{f_{\mathrm eq,i}(p')(1-f_{\mathrm eq,i}(p'))\tau_i}{T}\right)
\left\{\frac{p^0}{T}\partial_{0}T 
+T\partial_0\left(\frac{\mu}{T}\right)\right.\right.\nonumber\\
 &&\left. \left. -\frac{p^jp^{\nu}}{p^0} \partial_{j}u_{\nu}
-2q_i\frac{{\bf E \cdot p}}{p^0}\right\} \right].
\label{heatflow_temp}
\end{eqnarray}
In order to define the thermal conductivity for a system,
the number of particles in that system must be conserved
and therefore it requires the associated chemical potential to
be nonzero. In the ultrarelativistic heavy-ion collisions at RHIC
and LHC, the value of chemical potential $(\mu)$ is very small 
\cite{Munzinger:JPG28'2002,Cleymans:JPG35'2008,Andronicet:NPA837'2010}, 
its value extracted from the charged particle 
ratios are in the range of 50-100 MeV. In Navier-Stokes equation, 
the heat flow vector 
is related to the gradient of the thermal potential, $U_i=\mu_i/T$ as
\begin{eqnarray}
Q^i_{\mu}&\stackrel{\rm Def}=&-\kappa \frac{n_{\mathrm eq,i}}{\epsilon_i 
+P_i} T^2 \nabla_{\mu}U_i,\nonumber\\
&=&\kappa\left[\nabla_{\mu}T-\frac{T}{\epsilon_i + P_i}\nabla_{\mu}P_i
\right],
\label{Navier_stokes}
\end{eqnarray}
where the coefficient, $\kappa$ is the thermal conductivity and 
$\nabla_{\mu}=\partial_{\mu}-u_{\mu}u_{\nu}\partial^{\nu}$.
In the local rest frame of the fluid, only the spatial ($j$) component
of the heat flow four vector for $i$-th species is retained and takes the form 
\begin{eqnarray}
{Q^i_j} &=&-\kappa\left[\frac{\partial T}{\partial x_j}
-\frac{T}{n_{\mathrm eq,i}h_i}\frac{\partial P_i}{\partial {x_j}}\right].
\label{heat_navier1}
\end{eqnarray}

\begin{center}
\begin{figure}
\begin{tabular} {c c}
\includegraphics[width=6cm]{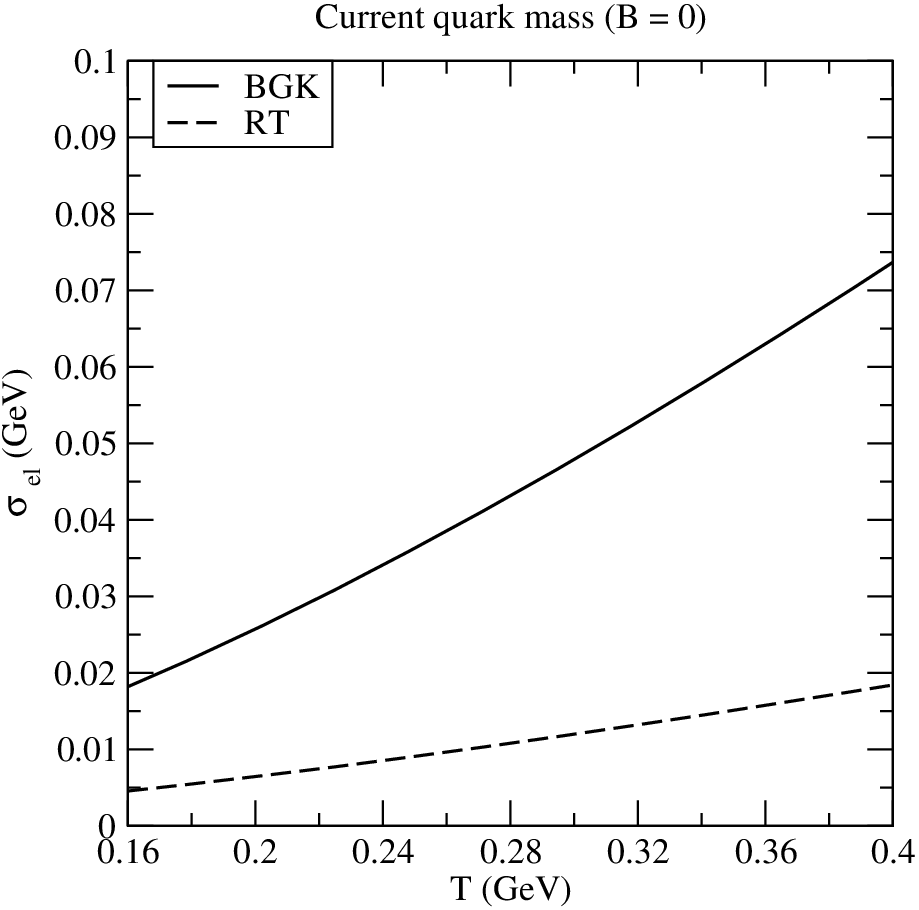}&
\hspace{1cm}
\includegraphics[width=6cm]{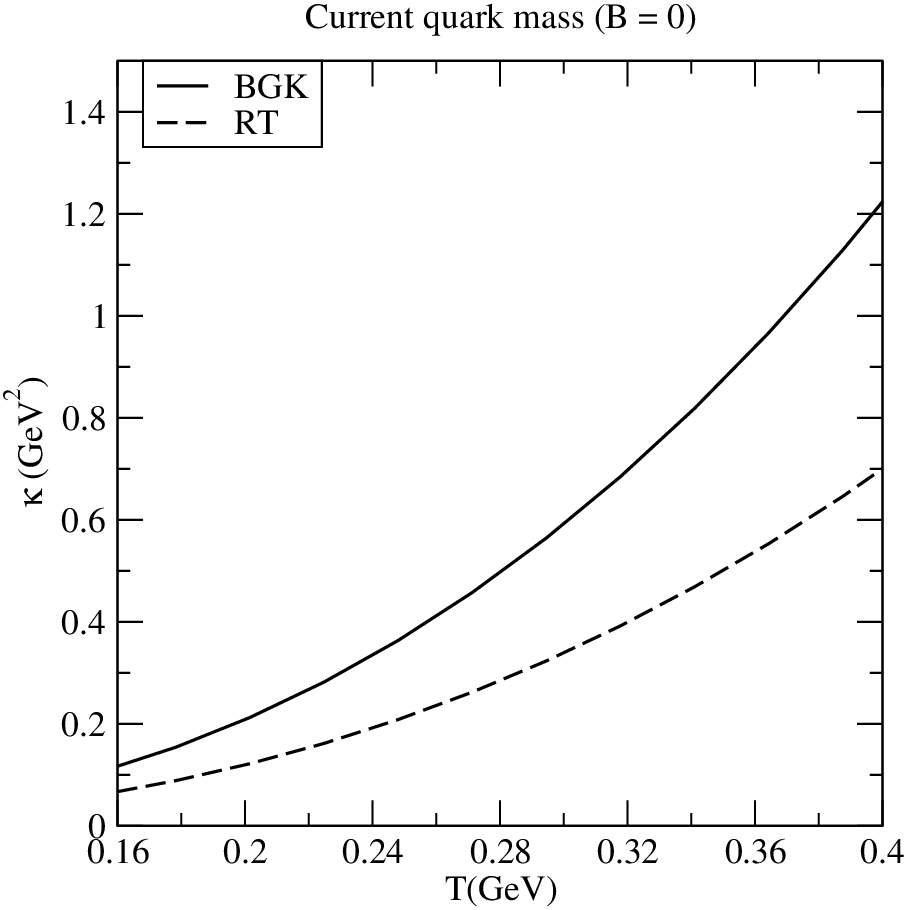}\\
(a) & (b)
\end{tabular} 
\caption {Variation of  electrical ($\sigma_{el}$) (left) and  
thermal ($\kappa$) (right) conductivities with the temperature.} 
\label{fig1}
\end{figure}
\end{center}

Thus, we have obtained the thermal conductivity by comparing the heat
flow (${\mathbf Q}$) calculated from the kinetic theory~\eqref{heatflow_temp} 
with its definition~\eqref{heat_navier1}. The effect of the modified 
collision term on the thermal
conductivity can be decomposed in terms of the contribution
due to the collision term of the relaxation type (RT) and a correction term
as,
 \ba \label{kappa_bgk} 
\kappa &=& \kappa^{\rm RT}+ \kappa^{\rm Corr},
\ea
where
\ba  
\kappa^{\rm RT} &=& \beta^2 \sum_i 2g_i \tau_i \left[\int_p
\frac{p^2}{3\omega_i^2(p)}(\omega_i(p)-h_i)^2f_{\rm eq,i}(p)
(1-f_{\rm eq,i}(p))\right].\\
\kappa^{\rm Corr} &=& \beta^2 \sum_i 2g^2_i \tau_i n_{\mathrm eq,i}^{-1}
\int_p \frac{p}{\omega_i(p)}(\omega_i(p)-h_i)f_{\rm eq,i}(p)\nonumber\\
&& \times \int_{p^\prime} \frac{p'}{\omega_i(p')}
(\omega_i(p')-h_i)  f_{\rm eq,i}(p')(1-f_{\rm eq,i}(p')).
\label{kappa_corr}
\ea
Thus, to visualize the effect of collision term on the charge and heat 
transport, we have computed both conductivities as a function of temperature
in both scenario of collision integrals in Figure \ref{fig1}, wherein we have
considered $u$, $d$, $s$ flavours with their current masses.
We have found that the modified BGK collision term enhances both charge and 
heat transport, compared to the collision term of relaxation type.
To be specific, the ratio of $\sigma_{\rm el}$ in modified BGK collision
term to the relaxation collision term is approximately 4.0, whereas the ratio
for $\kappa$ is $\sim$ 1.76, implying that the collision integral
is more sensitive to the charge transport.


\subsection{Wiedemann-Franz law and Knudsen number in the absence 
of magnetic field}
Wiedemann-Franz Law indicates that the transport of the 
charge and heat \footnote{{\em at least,} by the charged
particles alone} are not entirely different, {\em rather} the ratio of the
transport coefficients in the respective cases is proportional to the 
temperature
\begin{eqnarray}
\frac{\kappa}{\sigma_{el}}=L_RT,
\end{eqnarray}
and the proportional factor, $L_R$ is known as Lorentz number. The metals, 
which are good conductors 
of both electricity and heat, expectedly obey the Wiedemann-Franz law 
perfectly. There are some calculations in which the statement 
of the Wiedemann-Franz law have been violated, such as for the
strongly-coupled QGP medium \cite{Mitra:PRD96'2017},
two-flavor quark matter in the NJL model \cite{Harutyunyan:PRD95'2017}, 
and thermally populated electron-hole plasma in graphene,
describing the signature of a Dirac fluid \cite{Crossno:Scince351'2016}.
\begin{center}
\begin{figure}[H]
\begin{tabular}{cc}
\includegraphics[width=6cm]{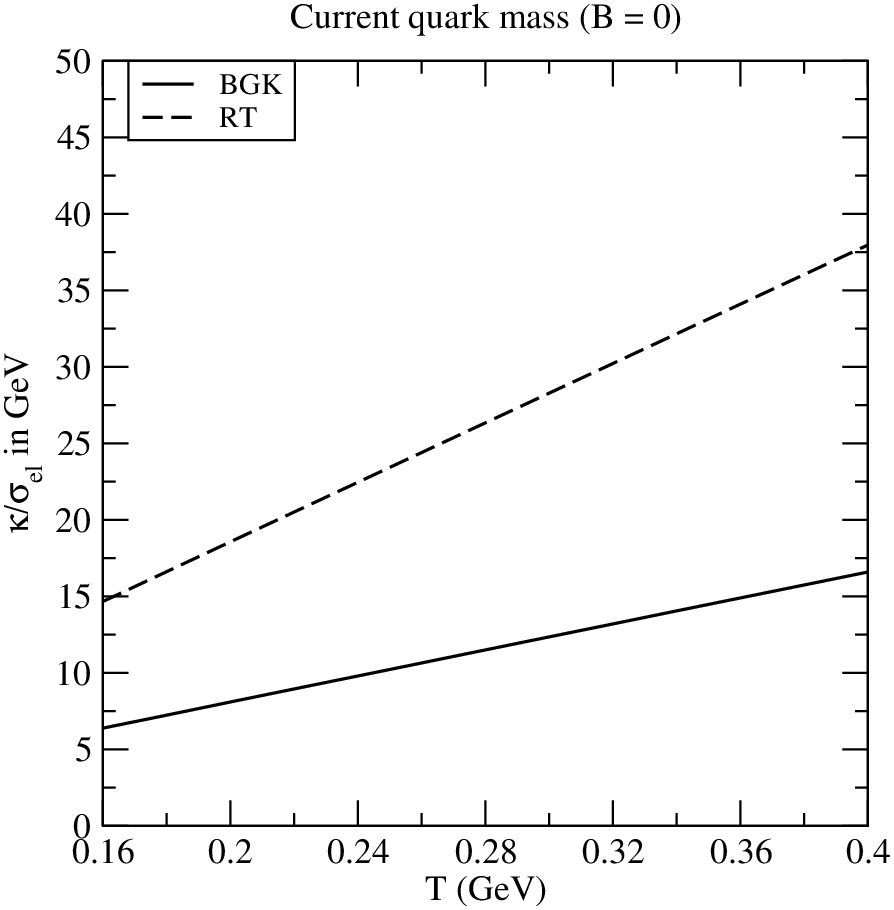}&
\hspace{1cm}
\includegraphics[width=6cm]{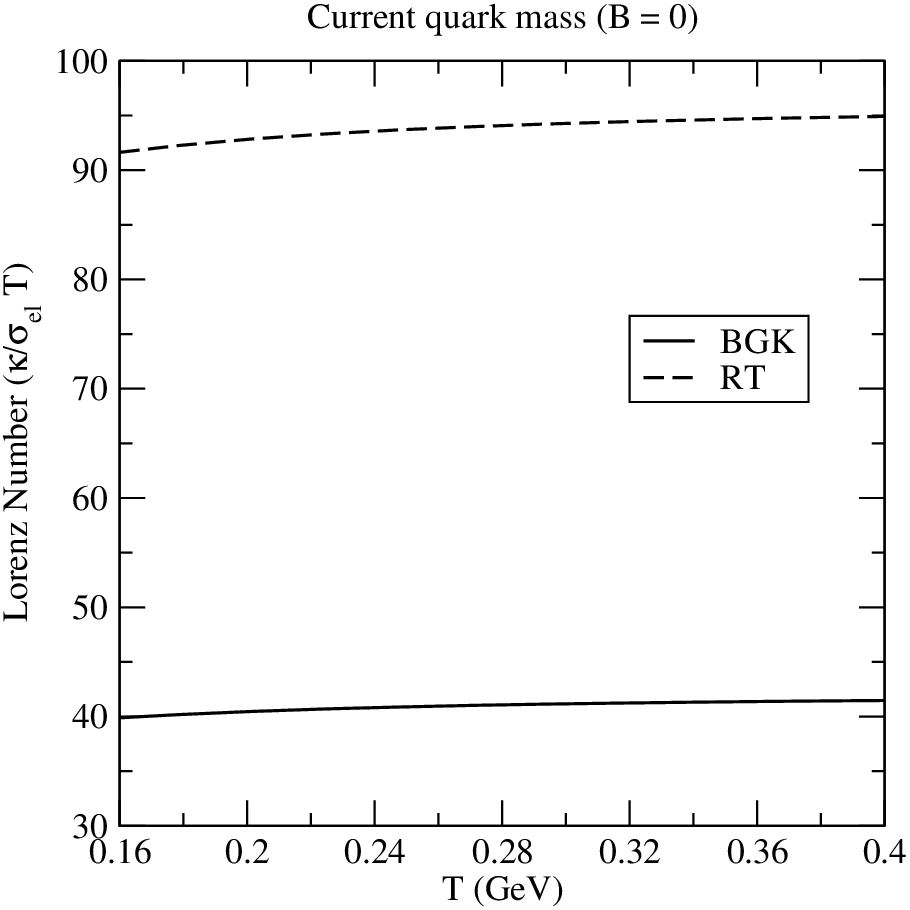}\\
(a)&(b)
\end{tabular}
\caption{Variation of the ratio of the thermal conductivity to the electrical conductivity  ($\kappa/\sigma_{el}$) and Lorenz number 
with temperature.} 
\label{fig2} 
\end{figure}
\end{center}
So far we have discussed the transport of both heat and electricity due to the
participation of quarks only, thus it becomes reasonable to think
that the heat and charge transport are not mutually exclusive as per
the statement of Wiedemann-Franz law. In fact, they are found 
to obey the aforesaid law, as seen in Figure \ref{fig2}(a). Although
the ratio, $\kappa/\sigma_{\rm el}$ is found to vary almost linearly with the
temperature in Figure \ref{fig2}(a), but the actual behaviour
of the interplay of heat and charge transport can be better understood
through the Lorenz number, $L_R$ (=$\kappa/(\sigma_{\rm el} T)$).
The Lorenz number initially increases monotonically in relatively
smaller temperature and behaves constant in the high temperature, as seen 
in Figure \ref{fig2}(b). The effect of collision terms seen in the conductivities (in 
Figure \ref{fig1}) get also reflected into the relative behaviour 
through the ratio, $\kappa/\sigma_{\rm el}$ and in the Lorenz number, $L_R$
as well, where the  relaxation collision integral is found to dominate over
the modified BGK collision integral.

The validity of the assumption of a system to be in local equilibrium 
is tested with the help of the Knudsen number ($\Omega$). It is defined 
as the ratio of the mean free path $(\lambda)$ to the characteristic 
length scale $(L)$ of the system as
\begin{eqnarray}
\Omega=\frac{\lambda}{L}.
\end{eqnarray}
The mean free path, $\lambda$ is computed from the 
thermal conductivity of the medium as
\ba
\lambda=\frac{3\kappa}{{\rm v_{rel}}C_V},
\ea
where $C_V$ is the specific heat of the medium and ${\rm v_{rel}}$
is the relative speed. The specific heat is contributed both 
by quarks and gluons, 
\ba
C_V=C_V^q+C_V^g,
\ea
where the quark contribution is 
\ba
C_V^q &=& \frac{\partial \epsilon^q}{\partial T}, \nonumber\\
&=&\frac{\beta^2}{\pi^2}\sum g_i\int dp~
p^2\omega_i^2 f_{\rm eq,i}(1-f_{\rm eq,i}),
\ea
and the gluon contribution is 
\ba
C_V^g &=& \frac{\partial \epsilon^g}{\partial T}, \nonumber\\
&=&\frac{\beta^2 g_g}{2\pi^2}\int dp~ p^2\omega_g^2 f_g (1+f_g),
\ea
where $f_g$ is the gluon distribution function, given by
 \begin{eqnarray}
 f_g=\frac{1}{e^{\beta \omega_g}-1}.
 \label{dis_gluon} 
 \end{eqnarray}

\begin{center}
\begin{figure}
\begin{tabular}{cc}
\includegraphics[width=6cm]{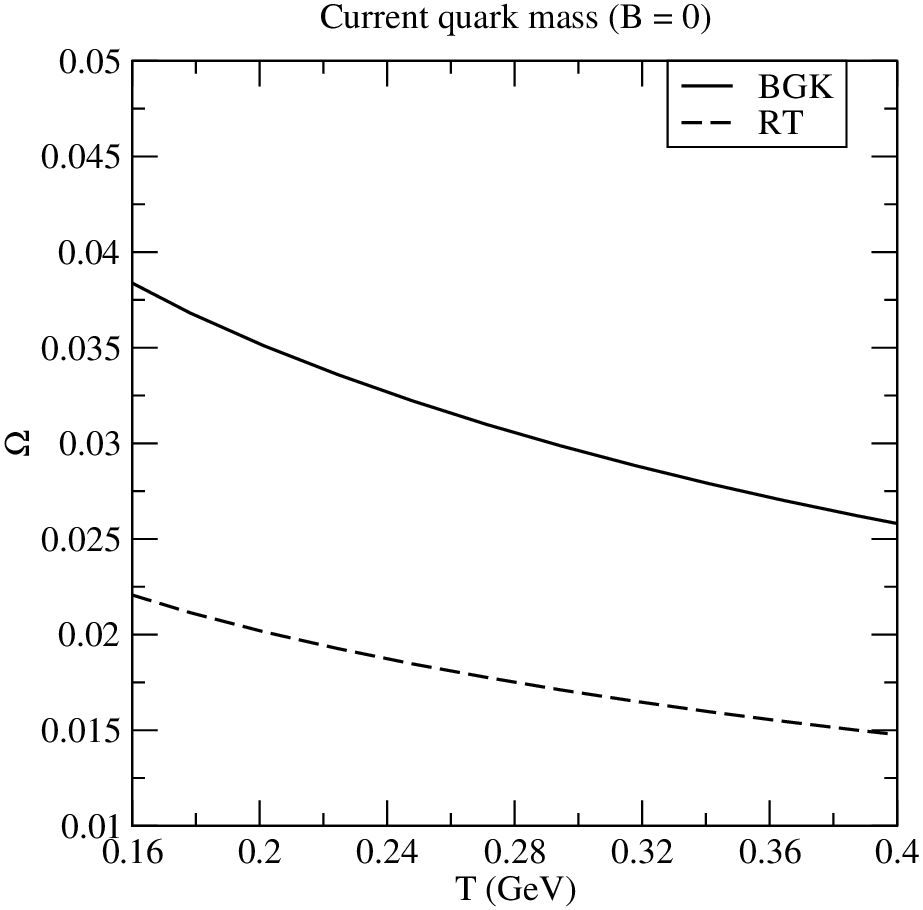}&
\hspace{1cm}
\includegraphics[width=6cm]{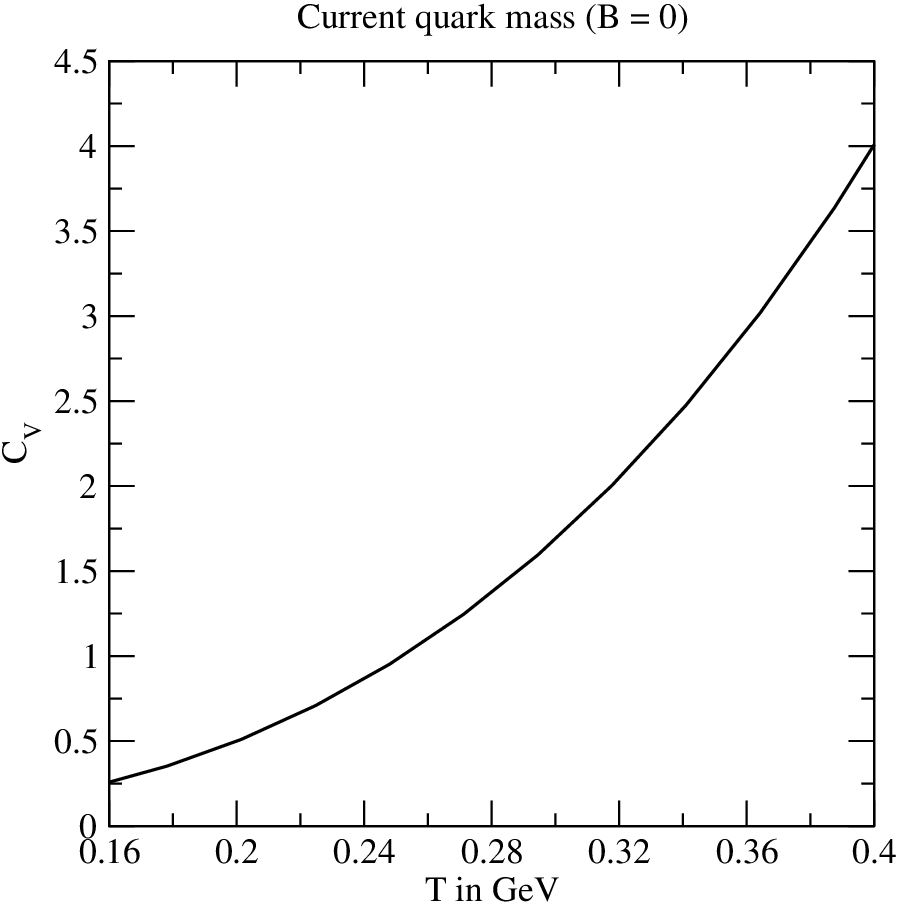}\\
(a)&(b)
\end{tabular}
\caption{Variation of the Knudsen number $(\Omega)$ and specific heat
with temperature.}
\label{fig3}  
\end{figure}
\end{center}

Finally, we have computed the Knudsen number as a function of
temperature in Figure \ref{fig3}(a), wherein the modified BGK collision 
term is found to dominate over its counterpart (RT) and can be understood
from the behaviour of $\kappa$. The magnitude of $\Omega$ is seen 
much lesser than one and decreases with the temperature, thus ensures
the validity of the system being in local equilibrium. This can be understood 
by the competition between $C_V$, which is measure of the heat content, 
and $\kappa$, a measure of the ease by which the heat can be transported 
in the system. $\Omega$  becomes
smaller than one because both the magnitude and the rate of increase of 
$C_V$ with $T$ is greater than the same in $\kappa$ (seen in Figure\ref{fig3}(b)). 

Thus, the sections 2.1 and 2.2 decipher the effect of the modified BGK 
collision term on the transport coefficients of the charge and heat, which, 
in turn, do the same for the Lorenz and Kundsen number in section 2.3.
In next section(s) 2.4 and 2.5, we will see how the strong magnetic 
field modulate the effect of collision term on the charge and heat transport,
respectively. This, in turn, will explore the same on the Lorenz and Knudsen 
number in Section 2.6.

\subsection{Electrical conductivity in a background of strong 
magnetic field}
In the presence of the strong magnetic field ($B$), the motion of quarks
becomes purely longitudinal (in the direction of magnetic field, {\em say}
3-direction), which is evident from the quantum mechanical relation: 
$\omega_i= \sqrt{p_3^2+m_i^2}$ in strong $B$. This is because
that only the lowest Landau level ($n=0$) is populated due to the
larger value of $|q_iB|$ ($\gg T^2$). Thus, when the medium is perturbed 
by the external magnetic field in 3-direction, an electromagnetic current 
is generated only in the longitudinal (3-) direction,
\begin{eqnarray}
J_3= 2\sum_i q_ig_i \frac{|q_iB|}{4\pi^2}\int {dp_3}~\frac{p_3}{\omega_i} 
~\delta f^{B}_i(x,p).
\label{current1}
\end{eqnarray}
Thus, the electrical conductivity relates the longitudinal current generated 
to the electrical field as
\begin{eqnarray}
J_3=\sigma_{el}~E_3,
\label{current2}
\end{eqnarray}
where $\sigma_{\rm el}$ is the longitudinal component of the 
electrical conductivity of the medium and the term `longitudinal' refers
with respect to the direction of the strong magnetic field. The component
of $\sigma_{\rm el}$ transverse to the magnetic field vanishes due to Landau quantization of the 
transverse motion in  the strong magnetic field\cite{fukushima:PRL120'2018}.

Therefore, in order to obtain $\sigma_{\rm el}$, we need to know the 
infinitesimal deviation of the medium ($\delta f^{B}_i$) in the presence 
of strong magnetic field. As earlier, this will be obtained by solving the 
Boltzmann equation \eqref{rbte} with the linearized collision 
integral. So, we start with the Boltzmann equation in a strong magnetic field
as
\begin{eqnarray}
p^0\frac{\partial f^B_i}{\partial x^0}+p^3\frac{\partial f^B_i}
{\partial x^3}+q_i F^{03}p_3\frac{\partial f^B_i}{\partial p^0}
+q_iF^{30}p_0\frac{\partial f^B_i}{\partial p^3}=-p^{\mu}u_{\mu}\nu_i^B 
\left(f_i^B-n_i^B{n_{\mathrm eq,i}^{B~-1}} f_{\mathrm eq,i}^B \right),
\label{RBTE_sigma}
\end{eqnarray}
where $n_i^B$ is the perturbed density and $n_{\mathrm eq,i}^B$ is
the equilibrium density in the presence of the strong magnetic field,
which is given by
\ba
n_{\mathrm eq,i}^{B}&=&\frac{g_i|q_iB|}{4\pi^2}\int dp_3 ~f_{\rm eq,i}^B,
\ea
with the equilibrium distribution function 
\begin{eqnarray}
f_{\mathrm eq,i}^{B}=\frac{1}{e^{\beta \omega_i}+1}.
\end{eqnarray}
The collision frequency in a strong $B$, $\nu_i^B$ runs with 
the (longitudinal) momentum ($p_3$), {\em unlike} that $\tau_i$
in pure thermal medium (\eqref{tau_B0}) is independent of momentum.
Moreover, $\tau_i^B$ depends strongly on the magnetic field and weakly
on the temperature because the dominant scale assigned to quark degrees of 
freedom in a thermal medium in strong $B$, being the magnetic field, in the 
same way that the 
temperature dominates in a thermal medium in the absence of magnetic field. 
The collision time, the inverse of the collisional frequency
($\nu_i^B$), in strong $B$ has been calculated recently 
in~\cite{Hattori:PRD95'2017}
\begin{eqnarray}
\tau_i^B(p_3;T,|q_iB|) = \frac{\omega_i (e^{\beta\omega_i}-1)}
{\alpha_s C_Fm_i^2(e^{\beta\omega_i}+1)}
{\left(\int \frac{dp'_3}{\omega'_ i
(e^{\beta \omega'_i}+1)}\right)}^{-1},
\end{eqnarray}
where $C_F$ (=4/3) is the Casimir factor and the strong
coupling, $\alpha_s$ now runs with the magnetic field
only
 \begin{eqnarray}
\alpha_s(|q_fB|)=\frac{1}{(\alpha^0(\mu_0))
^{-1}+\frac{11N_C}{12\pi}
\ln \left(\frac{{\rm k_z}^2+M_B^2}{\mu_0^2}\right)+\frac{1}{3\pi}
\sum_i \frac{|q_iB|}{\sigma}},
\label{coupling_B}
\end{eqnarray}
where  $$\alpha^0(\mu_0)=\frac{12\pi}{11N_C \ln
\left(\frac{\mu_0^2+M_B^2}{\Lambda_V^2}\right)}.$$
Here $M_B$ is infrared mass (1 GeV) and $\Lambda_V$ and $\mu_0$
are taken as 0.385 GeV and 1.1 GeV, respectively and ${\rm k_z}
=0.1\sqrt{eB}$.

Thus, going to the next approximation beyond the initial 
(equilibrium) configuration,  the total time-derivative
of the probability distribution in Boltzmann equation in the presence of 
strong magnetic field is simplified into
\begin{eqnarray}
\frac{df^B_i}{dt}= -2q_i\beta p_3 E_3 f_{\mathrm eq,i}^{B}\left( 
1-f_{\mathrm eq,i}^{B} \right),
\end{eqnarray}
wherein the partial derivatives of the equilibrium distribution, 
$f_{\mathrm eq,i}^B$ have been used:
\begin{eqnarray}
\label{partial2}
\frac{\partial f_{\mathrm eq,i}^{B}}{\partial p^0}&=&-\frac{1}{T} 
f_{\mathrm eq,i}^{B}(1- f_{\mathrm eq,i}^{B}),\\
\label{partial3}
\frac{\partial f_{\mathrm eq,i}^{B}}{\partial p^3}&=&-\frac{p_3}{Tp_0}
 f_{\rm eq,i}^{B}(1-f_{\rm eq,i}^B).
\end{eqnarray}
Next, linearizing the collision integral, the 
Boltzmann equation gives the transcendental equation for the
linear infinitesimal disturbance, $\delta f_i^B$ for $i$-th flavour as
\footnote{Using the symbol for one-dimension momentum
integration in strong $B$, $\int_{p_3} =\frac{|q_iB|}{2\pi}\int \frac{{dp_3}}{2\pi} $.}
\begin{eqnarray}
\delta f_i^B- g_i{n_{\mathrm eq,i}^{B}}^{-1} f_{\rm eq,i}^B \int_{p_3}\delta f_i^B
=2q_i\beta \tau_i^B \frac{p_3 E_3}{\omega_i} f_{\rm eq,i}^{B}\left( 
1-f_{\rm eq,i}^{B} \right),
\label{deviation}
\end{eqnarray}
 which yields the disturbance, $\delta f_i^B$ up to the first-order 
(neglecting ${\cal O} {(\delta f_i^B)}^2$ terms) as
\begin{eqnarray}
\delta f_i^B=\delta f_i^{B(0)}+ g_i{n_{\mathrm eq,i}^{B}}^{-1} 
f_{\rm eq,i}^B \int_{p'_{3}}\delta f_i^{B(0)},
\label{eqoneseven}
\end{eqnarray}
where
\begin{eqnarray}
\delta f_i^{B(0)}=\frac{2q_i\beta \tau_i^B}
{\omega_i} p_3 E_3~f_{\rm eq,i}^{B}(1-f_{\rm eq,i}^{B}).
\end{eqnarray} 
  
Therefore, we can now calculate the current density, $J_3$ from \eqref{current1}
with $\delta f_i^{B}$ taken from \eqref{eqoneseven} as
\ba
J_3&=&\frac{\beta}{\pi^2} \sum_i q_i^2 g_i
 |q_iB|\left[\int {dp_3}\frac{p_3^2}{\omega_i^2}~ \tau_i^B(p_3)~ 
 f_{\rm eq,i}^{B}(1-f_{\rm eq,i}^{B})\right. \nonumber \\
&& \left. +g_i{n_{\mathrm eq,i}^{B}}^{-1} \int {dp_3}\frac{p_3}{\omega_i}f_{\rm eq,i}^B
\int_{p'_{3}}\frac{p'_3}{\omega_i}~\tau_i^B (p_3^\prime)~f_{\rm eq,i}^{B}~(1-f_{\rm eq,i}^{B})\right]E_3   .
\ea

So we extract the coefficient of the electric field, $E_3$ 
as the electrical conductivity of a thermal QCD medium  in a 
strong $B$, which, in modified BGK collision term, yields 
as the sum of the contribution due to the RT collision term 
and a correction term 
\begin{eqnarray}
\sigma_{\rm el}^B = \sigma_{\rm el}^{B,\rm RT}+\sigma^{B,\rm Corr}_{el},
\label{sigma_cond}
\end{eqnarray}
where 
\ba
\sigma^{B,\rm RT}_{el} &=& \frac{\beta}{\pi^2} \sum_i q_i^2 g_i
 |q_iB|\int {dp_3}\frac{p_3^2 ~\tau_i^{B}}{\omega_i^2 } 
 f_{\mathrm eq,i}^{B}(1-f_{\mathrm eq,i}^{B}),\\
\sigma^{B,\rm Corr}_{el} &=& \frac{\beta}{\pi^2} \sum_i q_i^2 g^2_i |q_iB|
{n_{\mathrm eq,i}^B}^{-1} \left[ \int {dp_3}\frac{p_3}{\omega_i} f_{\rm eq,i}^B
\int_{p'_{3}}\frac{p'_3}{\omega_i}~\tau_i^B (p_3^\prime)~f_{\rm eq,i}^{B}~
(1-f_{\rm eq,i}^{B})\right],
\label{sigma_corr} 
\ea
where the correction, $\sigma^{\rm B,Corr}_{\rm el}$ is found to be 
positive, implying that even in the presence of strong magnetic field, the 
dominance of modified BGK collision integral over RT Collision integral is
still retained in the charge transport. 
\subsection{Thermal conductivity in a strong magnetic field} 
In this section, we will calculate  the
thermal conductivity of a hot QCD medium in the presence of
strong magnetic field. Thus, we closely follow the previous section
2.2.
In the presence of the strong magnetic field ($B$), only the component along 
the direction of the magnetic field (3-direction) survives and takes the form
\begin{eqnarray}
Q_3=\sum_i \frac{g_i|q_iB|}{2\pi^2}\int {dp_3}~\frac{p_3}
{\omega_i}~(\omega_i-h_i^B)~\delta f_i^B,
\label{heat_vec}
\end{eqnarray}
where 
\ba
h_i^{B}&=&\frac{(\epsilon_i^{B}+P_i^{B})}{n_{eq,i}^{B}},\\
\epsilon_i^{B}&=&\frac{g_i|q_iB|}{4\pi^2}\int dp_3 ~\omega_i f_{\rm eq,i}^B,\\
P_i^{B}&=&\frac{g_i|q_iB|}{12\pi^2}\int dp_3 \frac{p_3^2}{\omega_i} ~f_{\rm eq,i}^B.
\ea
On the other hand, the spatial component of 
heat flow vector \eqref{Navier_stokes} in the Navier-Stokes equation, takes
the form in a strong $B$ (along the 3-direction), 
\begin{eqnarray}
Q^i_3 &=&-\kappa\left[\frac{\partial T}{\partial x_3}
-\frac{T}{n^B_{eq,i}h^B_i}\frac{\partial P_i}{\partial x_3}\right],\nonumber\\
&=& \kappa\left[{\partial_3 T}
-\frac{T}{n^B_{\mathrm eq,i}h^B_i}{\partial_3 P_i}\right],
\label{heat_navier}
\end{eqnarray}
where the coefficient, $\kappa$ is the thermal conductivity.
Therefore, once we compute the heat flow vector from the
kinetic theory \eqref{heat_vec} and
express it in the form \eqref{heat_navier}, we could then pick up
the coefficient of gradient term as $\kappa$.
So we start with the Boltzman equation in the presence of strong 
magnetic field, in terms of velocity and temperature gradients through
the chain rule of differentiation, 
\begin{eqnarray}
p^{\mu} \frac{\partial f^B_i}{\partial T}
\frac{\partial T}{\partial x^{\mu}}
+p^{\mu} \frac{\partial f^B_i}{\partial p^0}
\frac{\partial (p^{\nu} u_{\nu})}{\partial x^{\mu}}
+q_i\left[F^{03}p_3
\frac{\partial f^B_i}{\partial p^0}+F^{30}p_0\frac{\partial f^B_i}{\partial p^3}
\right]=C[f].
\label{RBTE_heat}
\end{eqnarray}
For vanishingly small value of $\mu_i$, the 
infinitesimal deviation, $\delta f_i^B$ is obtained by solving the 
Boltzman equation, after linearizing the collision integral
with respect to the deviation { (see the Appendix B)}  
\begin{eqnarray}
\delta f_{i}^{B}=\delta f_i^{B(1)}+ g_i{n_{\mathrm eq,i}^{B}}^{-1}f_{\rm eq,i}^B
\int_{p'_{3}} \delta f_i^{B(1)},
\end{eqnarray}
where 
\begin{eqnarray}
\delta f_i^{B(1)}&=&-\frac{f_{\rm eq,i}^{B}(1-f_{\rm eq,i}^{B})\tau_i^B}{T}
\left[\frac{p^0}{T}\partial_{0}T+\frac{(p^0-h_i^B)}
{p^0}\frac{p^3}{T}\left(\partial_3 T-\frac{T}{n^B_{\mathrm eq,i} h_i^B}
\partial_3 P_i\right)
+T\partial_0\left(\frac{\mu_i}{T}\right)\right.\nonumber \\
&&\left. -\frac{p^3p^{\nu}}{p^0} \partial_{3}u_{\nu}
-2q_i\frac{E_3p_3}{p^0}\right].
\end{eqnarray}

Now we are in a position to calculate the heat-flow vector ($Q_3$) 
in the presence of strong magnetic field from the kinetic theory
by plugging $\delta f_i^B$ in \eqref{heat_vec},
\begin{eqnarray}
Q_3 &=& \sum_i \frac{g_i|q_iB|}{2\pi^2}\int {dp_3}~\frac{p_3}
{\omega_i}~(\omega_i-h_i^B)\left[\left\{\frac{f_{\rm eq,i}^{B}
(1-f_{\rm eq,i}^{B})\tau_i^{B}}{T}
\frac{(p^0-h_i^B)}{p^0}\frac{p_3}{T}\right. \right. \nonumber\\
&& \left. \left. +g_i{n_{\mathrm eq,i}^{B}}^{-1} f_{\rm eq,i}^B\int_{p'_{3}}
\frac{f_{\rm eq,i}^{B}(1-f_{\rm eq,i}^{B})\tau_i^{B}}{T}\frac{(p^0-h_i^B)}
{p^0}\frac{p_3}{T}\right\} \left(\partial_ 3 T-\frac{T}{n^B_{\mathrm eq,i}
h_i^B} \partial_3 P_i\right)\right.\nonumber \\ 
&& \left. +\left(\frac{f_{\rm eq,i}^{B}(1-f_{\rm eq,i}^{B})\tau_i^{B}}{T}
+ g_i{n_{\mathrm eq,i}^{B}}^{-1}f_{\rm eq,i}^B\int_{p'_{3}}
\frac{f_{\rm eq,i}^{B}(1-f_{\rm eq,i}^{B})\tau_i^{B}}{T}\right)
\left\{\frac{p^0}{T}\partial_{0}T 
+T\partial_0\left(\frac{\mu}{T}\right)\right.\right.\nonumber\\
 &&\left. \left. -\frac{p^3p^{\nu}}{p^0} \partial_{3}u_{\nu}
-2q_i\frac{E_3p_3}{p^0}\right\} \right].
\label{heatflowcalculated}
\end{eqnarray}

Thus, we obtain $\kappa$ from the heat flow vector, by matching the 
coefficient of the gradient of thermal potential in the Navier-Stokes 
equation \eqref{heat_navier} with the same in
the kinetic theory \eqref{heatflowcalculated} and similar to the absence 
of (strong) magnetic field, $\kappa$ in modified collision term is 
decomposed into the contribution due to collision term of relaxation type 
and a correction term
\begin{eqnarray}
\mathbf{\kappa}^B= \kappa^{\rm B,RT} + \kappa^{\rm B,Corr},
\end{eqnarray}
where
\begin{eqnarray}
\kappa^{\rm B,RT} &=& \frac{\beta^2}{2\pi^2} \sum_i{g_i|q_iB|}
 \int {dp_3}\frac{p_3^2~\tau_i^{B}}
{\omega_i^2 }(\omega_i-h_i^B)^2f_{\rm eq,i}^{B}(1-f_{\rm eq,i}^{B}) 
\label{kappa_rt_B},\\
\kappa^{\rm B,Corr} &=& \frac{\beta^2}{2\pi^2} \sum_i{g^2_i|q_iB|}
{n_{\mathrm eq,i}^B}^{-1} \left[ \int {dp_3}\frac{p_3}
{\omega_i}(\omega_i-h_i^B)f_{\rm eq,i}^{B}\int_{p'_{3}}\frac{p'_3~\tau_i^{B}(p'_3)}
{\omega'_i}
(\omega'_i-h_i^B)f_{\rm eq,i}^{B}(1-f_{\rm eq,i}^{B})\right],
\label{kappa_B_corr} 
\end{eqnarray}
where the correction factor in strong $B$, unlike in the
absence of magnetic field \eqref{kappa_corr}, becomes negative. As a result, the 
dominance of the modified BGK collision term over the RT collision term 
is lost.

\begin{center}
\begin{figure}
\begin{tabular} {c c}
\includegraphics[width=6cm]{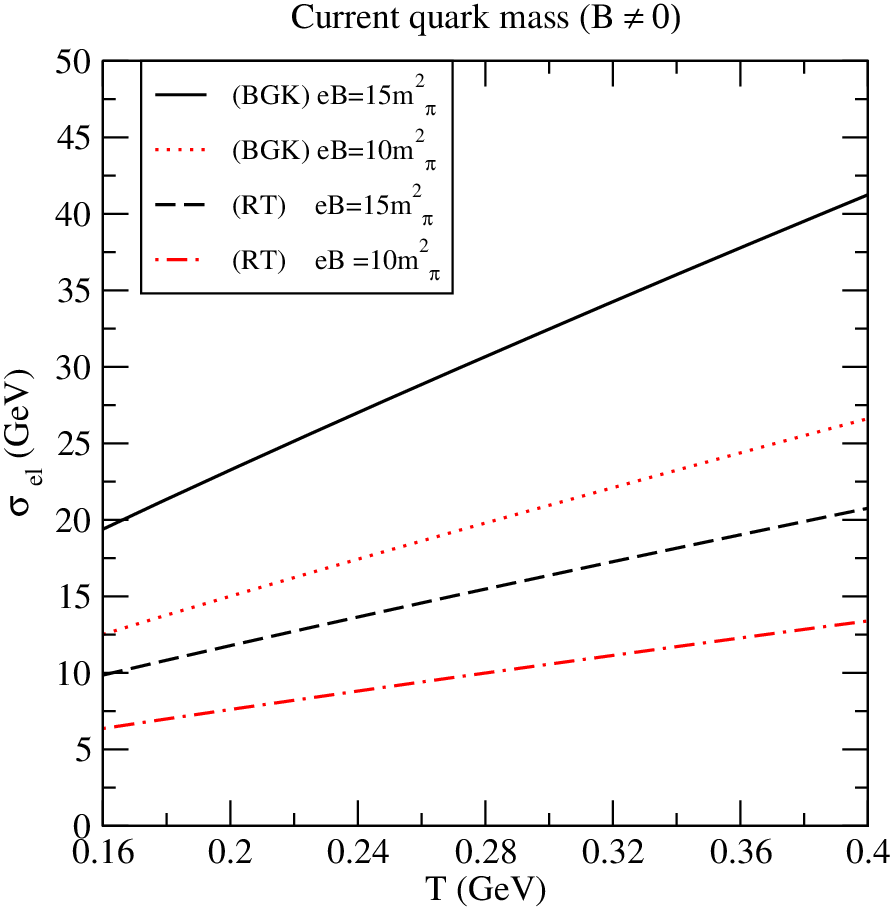}&
\hspace{1cm}
\includegraphics[width=6cm]{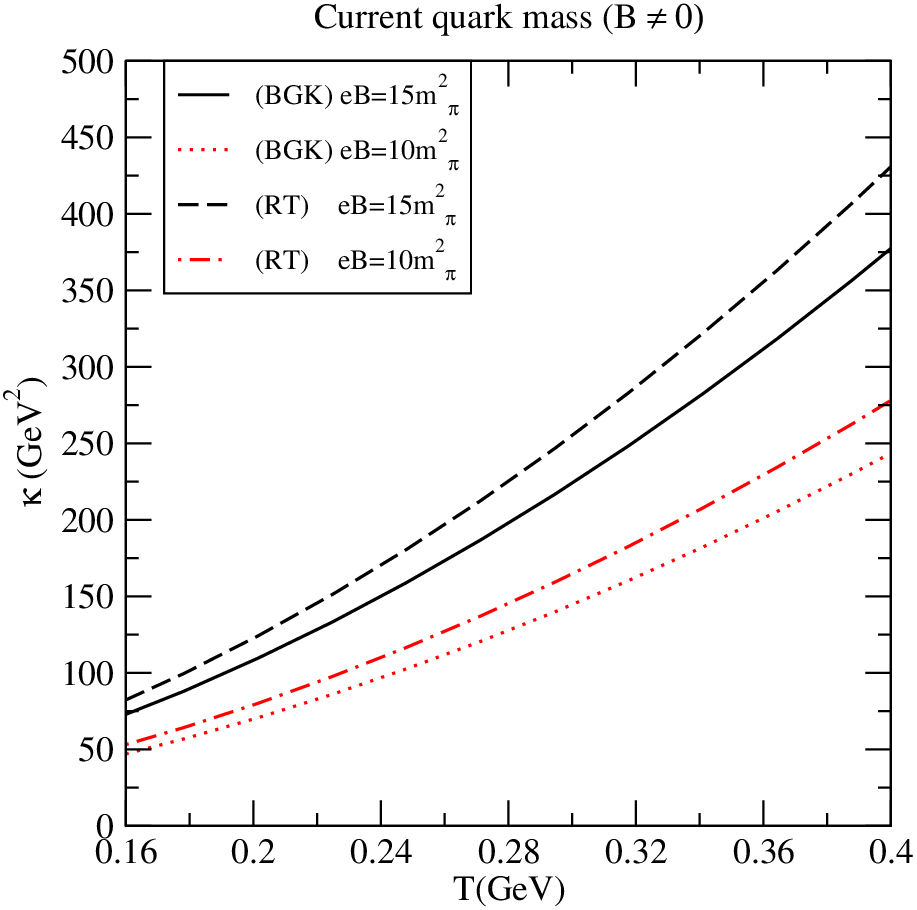}\\
(a) & (b)
\end{tabular} 
\caption {Variation of  electrical ($\sigma_{el}$) (left) and  
thermal ($\kappa$) (right) conductivities with the temperature.} 
\label{fig4}
\end{figure}
\end{center}
To see how the strong magnetic field
modulates the effect of collision term on the electrical and
thermal conductivities pictorially, we have computed them as a function
of temperature at increasing strengths of magnetic fields: $eB=10 m_\pi^2$ 
and $eB=15 m_\pi^2$. 
The observations are two folds: i) The strong $B$ 
enhances overall magnitudes of both $\sigma_{\rm el}$ and $\kappa$ 
by orders of two-to-three in high and low temperature region, respectively. 
Especially, the $B$ affects the charge transport with RT collision terms.
ii) it flips the dominance of the collision terms 
from BGK to RT in heat transport. This is due to the fact 
that {\em unlike in $B=0$ case}, the correction term in \eqref{kappa_B_corr}
becomes negative, because the dispersion relation in strong $B$
makes the relaxation-time momentum dependent\footnote{unlike the relaxation
time \eqref{tau_B0} in pure thermal medium (B=0)}, which in low 
momentum limit ($p_3 \rightarrow 0$) becomes constant and in high momentum
limit ($p_3 \rightarrow \infty$), increases indefinitely with $p_3$. As
a result, the second integral in the correction, $\kappa^{\rm B, Corr}$
becomes  positive and makes the overall correction negative (because the 
first integral - $\int_{p_3} \frac{p_3}{\omega_i} (\omega_i -h_i^B)
f^B_{\mathrm eq,i}$
is always negative)\footnote{In the absence of strong 
magnetic field, the product of the negative contributions coming from 
first and second integral make the correction term \eqref{kappa_corr}
positive.}.

  Thus, having deciphered the sole effects of collision term and 
strong magnetic field on the electrical and thermal conductivities
for a medium consisting of noninteracting/ideal partons, we 
will examine how the abovementioned affects
modulates the interplay between heat and charge transport
and the affiliated transport coefficient derived by
them. Thus, the relative behaviour is checked by the ratio, 
$\kappa/\sigma_{\rm el}$ (and the Lorenz number, $L_R$) and the 
derived coefficient is the equilibration factor, quantified
by the Knudsen number ($\Omega$). 
It is found that both the ratio, $\kappa/\sigma_{\rm el}$ and the 
Knudsen number are dominated by the RT 
collision term over the BGK term in a fixed (strong) $B$ whereas 
for a collision term, the effect of strong $B$ can be understood 
as follows:

One of the factor in the denominator of the Knudsen number is 
the specific heat and the quark contribution to the specific heat 
gets severely affected due to the dimensional reduction in strong $B$ as
\ba
C_V^q 
&=& \frac{\partial \varepsilon^q}{\partial T} \nonumber\\
&=&\frac{\partial(u_{\alpha}T^{\alpha\beta}u_{\beta})}{\partial T}\\
&=&\beta^2\sum_i \frac{g_i|q_iB|}{2\pi^2}\int dp_3~
\omega_i^2 f_{\rm eq,i}^{B}(1-f_{\rm eq,i}^{B}),
\ea
resulting an overall decrease in the specific heat. As a consequence,
the Knudsen number becomes too large to understand
physically because the validation of the equilibrium thermodynamics 
requires $\Omega$ to be less than one  
because the large Knudsen number contradicts with the basic idea 
of near-equilibrium assumption that is used to linearize the kinetic 
equation. This motivates us to treat the partons interacting because 
the interactions among the partons in a thermal medium generate the 
masses, which, in turn, behaves as an infrared cut-off in the 
crosssection of $gg \rightarrow gg$ scattering, responsible for 
bringing the system into equilibrium. Hence, the unusually larger
value of relaxation-time in a strong magnetic field becomes finite 
and causes the thermal conductivity smaller. This possible way-out
resonates with the idea that the more the constituents interact amongst
themselves the quicker the system achieves the equilibrium.
Thus, the necessity of quasiparticle description of partons gets 
motivated in the next section.
 
\section{Charge and heat transport with the quasiparticle 
description of partons}
The transport of charge and heat in a thermal QCD medium with the 
noninteracting quarks and gluons in strong $B$ hereinabove described 
yields unusually large values of electrical and thermal conductivities 
and the affiliated coefficients
(Lorenz number, Knudsen number) therein, which defy physical interpretation.
As we know, the thermal medium generically ascribes the masses to the
constituents, which motivates us for a quasiparticle description
of partons participating in the transport phenomena in this section.

\subsection{Quasiparticle description of partons}
 
The idea of quasiparticle description (QPD) of a parton in
a medium is to encrypt the
interaction of the given parton with the remaining partons
in terms of its in-medium mass, {\em known as}
quasiparticle masses or thermal masses\footnote{in addition to 
its current mass} and then treat these quasiparticles
as noninteracting particles. In a sense, at the scale 
of the quasiparticle mass, the independent (single)behaviour of partons 
ceases to exist and the collective behaviour of the medium sets in. 
There are some variants of quasiparticle description where the interactions 
amongst themselves are also taken into account. Different versions
of quasi-particle description exist in the literature based 
on different effective theories, such as thermodynamically consistent
 quasi particle model \cite{Bannur:JHEP09'2007}, Nambu-Jona-Lasinio (NJL) 
model and its extension Polyakov NJL (PNJL model \cite{Fukushima:PLB591'2004,
Ghosh:PRD73'2006,Abuki:PLB676'2009}, Gribov-Zwanziger 
quantization model \cite{Su:PRL114'2015,Florkowski:PRC94'2016}.
In this work we employ the medium generated masses for quarks and
gluons in the framework of perturbative QCD at finite temperature and/or
strong $B$ from the poles of resummed propagators calculated from 
the respective self-energies.

Let us start with a thermal QCD medium in the absence of magnetic field.
The quasiparticle masses of $i$-th flavour is written phenomenologically
as~\cite{Bannur:JHEP09'2007} 
\begin{eqnarray}
m_i^2=m_{i0}^2+\sqrt{2}m_{i0}m_{iT}+m_{iT}^2,
\label{para_massT}
\end{eqnarray}
where $m_{i,0}$ and $m_{i,T}$ are the current quark mass and thermally generated
mass respectively. The thermal mass for the quarks have been calculated in 
hard-thermal-loop perturbation theory with the temperature as the 
hardest scale \cite{Braaten:PRD45'1992,Peshier:PRD66'2002} as
\begin{eqnarray}
m_{iT}^2=\frac{g'^2T^2}{6},
\label{quarkmassT}
\end{eqnarray}
where $g'$ is the strong coupling which runs with the temperature   
(eqn. \eqref{coupling_T}). Similarly the gluons  also acquire
a thermal mass, which is also calculated as
\ba 
m_g^2=\frac{g'^2T^2}{6}\left(N_C+\frac{N_f}{2}\right).
\label{gluonmassT} 
\ea

Now, for thermal QCD medium in the presence of strong magnetic field, 
the form of thermal/quasiparticle mass can be generalized as
\begin{eqnarray}
m_i^2=m_{i0}^2+\sqrt{2}m_{i0}m_{iT,B}+m_{iT,B}^2,
\label{para_massTB}
\end{eqnarray}
where the thermal mass, $m_{iT,B}$ are obtained 
from the dispersion relation of the full quark propagator in strong
$B$, by solving the Dyson-Schwinger equation self-consistently: 
\begin{eqnarray}
S^{-1}(p_{\parallel})=\gamma^{\mu}p_{\parallel \mu}-\Sigma(p_{\parallel}).
\end{eqnarray}
The $\Sigma(p_{\parallel})$ in the above is the quark self-energy, which
needs to be evaluated at finite temperature in the presence of 
strong $B$. Up to one-loop, its expression is  given by
\begin{eqnarray}
\Sigma(p)=-\frac{4}{3} g^{2}i\int{\frac{d^4k}{(2\pi)^4}}
\left[\gamma_\mu {S(k)}\gamma_\nu {D^{\mu \nu} (p-k)}\right],
\label{quark_self}
\end{eqnarray}
where the QCD coupling, $g$ now runs with the magnetic field,
mentioned in\eqref{coupling_B}).

The quark propagator, $S(k)$ in an external magnetic
field is calculated \cite{Schwinger:PR82'1951} by the
Schwinger proper-time method in the momentum space in terms
of Laguerre polynomials,
\begin{equation}
iS_n (k)=\sum_n\frac{-id_n(\alpha)D+d^\prime_n(\alpha)
\bar{D}}{k_\|^2 -m^2_f+2n|q_fB|}
+i\frac{\gamma\cdot k_\bot}{k_\bot^2}~,
\label{prop_lag}
\end{equation}
where the label, $n (=0,1,2, \cdots)$ denotes the Landau levels. The
above symbols are defined as~\cite{Chyi:PRD62'2000}
\begin{eqnarray*}
&&D=(m_f+\gamma\cdot k_\|)+\gamma\cdot
k_\bot\frac{m_f^2-k_\|^2}{k_\|^2},\\
&& \bar{D}=\gamma_{1}\gamma_{2}(m_f+\gamma\cdot k_\|),\\
&& d_n(\alpha)=(-1)^n e^{-\alpha}C_n(2\alpha),\\
&& d^{'}_{n}(\alpha)=\frac{\partial d_n}{\partial\alpha},
\end{eqnarray*}
with the dimensionless variable, $\alpha$ = $k_\bot^2/|q_fB|$.
The $C_n$'s are expressed in terms of Laguerre polynomial ($L_n$)
\begin{eqnarray*}
C_n(2\alpha)&=&L_{n}(2\alpha)-L_{n-1}(2\alpha).
\end{eqnarray*}

In the strong magnetic limit, only the lowest Landau levels get
populated, so $S(k)$ is simplified into a form
 \ba\label{quark_prop}
 S(k)=ie^{-\frac{k^2_\perp}{|q_iB|}}\frac{\left(\gamma^0 k_0-\gamma^3 k_z+m_     i\right)}{k^2_\parallel-m^2_i}\left(1
 -\gamma^0\gamma^3\gamma^5\right)
 ,\ea
 where the four vectors are defined with the metric tensors:
 $g^{\mu\nu}_\perp={\rm{diag}}(0,-1,-1,0)$ and $g^{\mu\nu}_\parallel=
 {\rm{diag}}(1,0,0,-1)$,
 \begin{eqnarray*}
 && k_{\perp\mu}\equiv(0,k_x,k_y,0), ~~ k_{\parallel\mu}\equiv(k_0,0,0,k_z)
 ~.\end{eqnarray*}
 
The gluon propagator, $D^{\mu \nu}(p-k)$ is not affected
by the magnetic field, so its takes the form 
 \ba
 \label{g. propagator}
 D^{\mu \nu} (p-k)=\frac{ig^{\mu \nu}}{(p-k)^2}
  ~.
\ea

In imaginary-time formalism, the quark self-energy \eqref{quark_self}
in strong magnetic field can be simplified into 
(see the Appendix C)\cite{Rath:PRD100'2019}
 \begin{eqnarray}
\Sigma(p_\parallel)=\frac{g^2|q_iB|}{3\pi^2}\left[\frac{\pi T}{2m_i}-\ln(2)\right]\left[\frac{\gamma^0p_0}{p_\parallel^2}+\frac{\gamma^3p_z}{p_\parallel^2}+\frac{\gamma^0\gamma^5p_z}{p_\parallel^2}+
\frac{\gamma^3\gamma^5p_0}{p_\parallel^2}\right],
\end{eqnarray}
which can further be decomposed in the covariant form as
\cite{Ayala:PRD91'2015,Karmakar:PRD99'2019}
\begin{eqnarray}
\Sigma(p_{\parallel})=A\gamma^{\mu}u_{\mu}+B\gamma^{\mu}b_{\mu}
+C\gamma^{5}\gamma^{\mu}u_{\mu}+D\gamma^{5}\gamma^{\mu}b_{\mu},
\end{eqnarray}
 where $A,B,C$ and D are the form factors. $u^{\mu}(1,0,0,0)$ 
and $b^{\mu}(0,0,0,-1)$ represents 
the direction of the heat bath and magnetic field respectively.
These vectors are behind  the breaking of the Lorentz
and the rotational symmetry respectively.
The form factors can be calculated in LLL approximation as
\begin{eqnarray}
A&=&\frac{1}{4}{\rm Tr}[\Sigma\gamma^{\mu}u_{\mu}]=
\frac{g^2|q_iB|}{3\pi^2}\left[\frac{\pi T}{2m_i}-\ln{(2)}\right]
\frac{p_0}{p_{\parallel}^2},\\
B&=&-\frac{1}{4}{\rm Tr}[\Sigma\gamma^{\mu}b_{\mu}]=
\frac{g^2|q_iB|}{3\pi^2}\left[\frac{\pi T}{2m_i}-\ln{(2)}\right]
\frac{p_z}{p_{\parallel}^2},\\
C&=&\frac{1}{4}{\rm Tr}[\gamma^5 \Sigma\gamma^{\mu}u_{\mu}]=-
\frac{g^2|q_iB|}{3\pi^2}\left[\frac{\pi T}{2m_i}-\ln{(2)}\right]
\frac{p_z}{p_{\parallel}^2},\\
D&=&-\frac{1}{4}{\rm Tr}[\gamma^5\Sigma\gamma^{\mu}b_{\mu}]=-
\frac{g^2|q_iB|}{3\pi^2}\left[\frac{\pi T}{2m_i}-\ln{(2)}\right]
\frac{p_0}{p_{\parallel}^2},
\end{eqnarray}
 we get $C=-B$ and $D=-A$. In terms of chiral projection operators, 
 the quark self energy takes the form
\begin{eqnarray}
\Sigma(p_{\parallel})=P_R[(A+C)\gamma^{\mu}u_{\mu}+(B+D)\gamma^{\mu}b_{\mu}]P_L
+P_L[(A-C)\gamma^{\mu}u_{\mu}+(B-D)\gamma^{\mu}b_{\mu}]P_R,
\end{eqnarray}
\begin{eqnarray}
\Sigma(p_{\parallel})=P_R[(A-B)\gamma^{\mu}u_{\mu}+(B-A)\gamma^{\mu}b_{\mu}]P_L
+P_L[(A+B)\gamma^{\mu}u_{\mu}+(B+A)\gamma^{\mu}b_{\mu}]P_R,
\end{eqnarray}
where $P_R=(1+\gamma^{5})/2$ and $P_L=(1-\gamma^{5})/2$ are 
the right handed and left handed chiral projection operators 
respectively. The effective quark propagator in terms of the 
chiral projection operators can be written as 
\begin{eqnarray}
S^{-1}(p_{\parallel})=P_R\gamma^{\mu}X_{\mu}P_L+P_L\gamma^{\mu}Y_{\mu}P_R,
\end{eqnarray}
where 
\begin{eqnarray}
\gamma^{\mu}X_{\mu}&=&\gamma^{\mu}p_{\parallel \mu}-
(A-B)\gamma^{\mu}u_{\mu}-(B-A)\gamma^{\mu}b_{\mu},\\
\gamma^{\mu}Y_{\mu}&=&\gamma^{\mu}p_{\parallel \mu}-
(A+B)\gamma^{\mu}u_{\mu}-(B+A)\gamma^{\mu}b_{\mu}.
\end{eqnarray}
The effective propagator can be further written as
\begin{eqnarray}
S(p_{\parallel})=\frac{1}{2}\left[P_R\frac{\gamma^{\mu}Y_{\mu}}{Y^2/2}P_L
+P_L\frac{\gamma^{\mu}X_{\mu}}{X^2/2}P_R\right]
\end{eqnarray}
where
\begin{eqnarray}
&&\frac{X^2}{2}=X_1^2=\frac{1}{2}\left[p_0-(A-B)\right]^2-\frac{1}{2}\left[p_z+(B-A)\right]^2 ~, \\
&&\frac{Y^2}{2}=Y_1^2=\frac{1}{2}\left[p_0-(A+B)\right]^2-\frac{1}{2}\left[p_z+(B+A)\right]^2
~.\end{eqnarray}
We take static limit ($p_0 = 0,~p_z \rightarrow 0$)  
of either $X_1^2$ or $Y_1^2$ (which are equal in this
limit) to get thermal mass (squared) at finite
 temperature and strong magnetic field as
\begin{eqnarray}
m_{iT,B}^2=\frac{g^2|q_iB|}{3\pi^2}\left[\frac{\pi T}
{2m_i}-\ln{(2)}\right],
\label{massTB}
\end{eqnarray}
which depends on both the magnetic field and the 
temperature.

\subsection{Electrical and Thermal Conductivity in B=0: Wiedemann Franz Law and 
Knudsen Number}
In kinetic theory approach of evaluating the transport coefficients, 
the quasiparticle description (QPD) of partons encodes the interactions 
present in the medium through the dispersion relation, 
where the vacuum masses in noninteracting scenario get modified by the 
quasiparticle masses (which depend on the $T$ and $B$).
Thus, the occupation probability (distribution function) gets modified
through the dispersion relation, which, in turn, will affect the transport
process of heat, charge etc. In short, we follow the derivation of 
the conductivities and their derived coefficients in QPD scenario, 
in the same way as was done in the previous section, except that the
distribution function now involves the masses generated by the
medium. We have found that the forms of the conductivities as in
noninteracting scenario remain the same, so we compute the electrical and 
thermal conductivities as a function of temperature with the in-medium masses 
\eqref{quarkmassT} and \eqref{gluonmassT} for quarks and gluons 
in the respective distribution functions \eqref{dis_quark} and \eqref{dis_gluon},
respectively. It is found that both charge and heat

\begin{center}
\begin{figure}[H]
\begin{tabular} {c c}
\includegraphics[width=6cm]{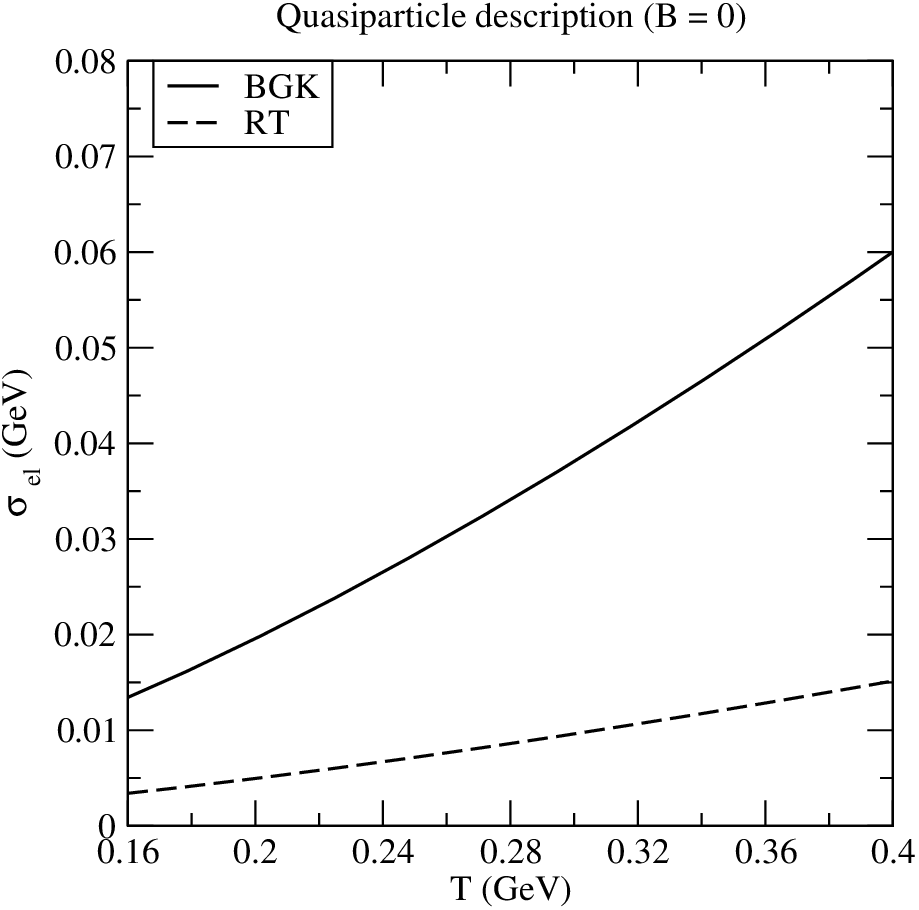}&
\hspace{1cm}
\includegraphics[width=6cm]{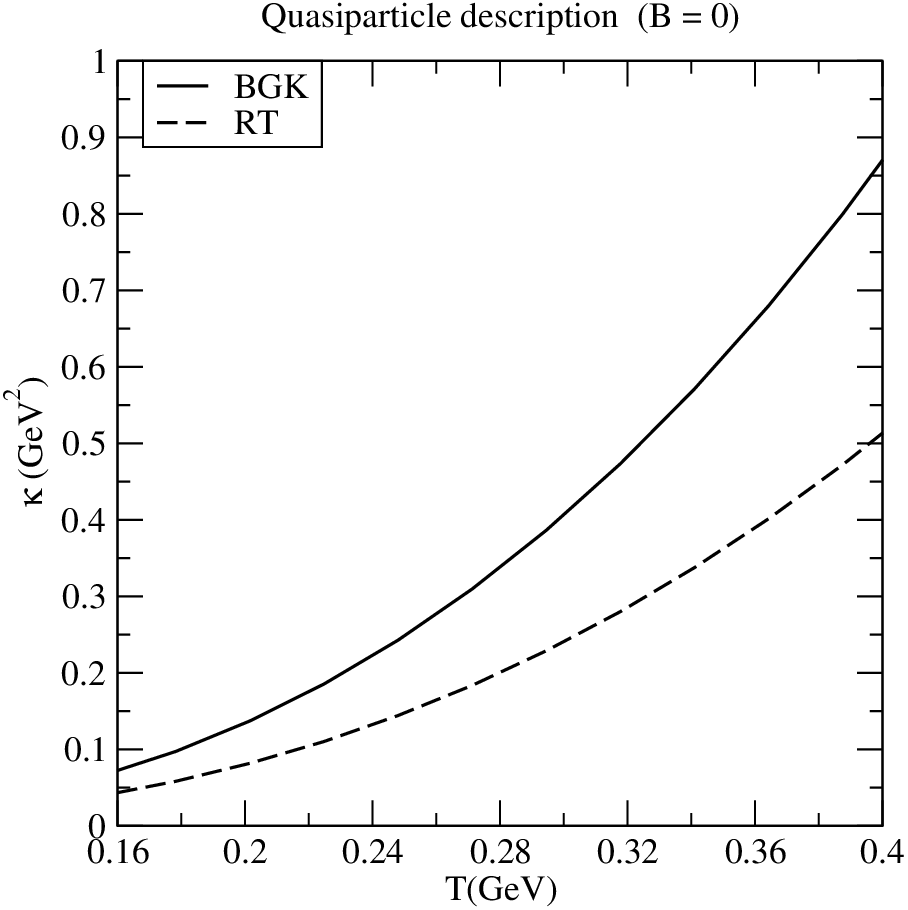}\\
(a)&(b)
\end{tabular}
\caption { Variation of  electrical conductivity $\sigma_{el}$ (left) and 
thermal conductivity $\kappa$ (right) with respect to the temperature } 
\label{fig7}  
\end{figure}
\end{center}

transport get impeded due to the in-medium (heavier) masses of quarks
(which reduce the mobility of the carriers), 
compared to the noninteracting scenario with current quark masses, 
which are reflected in the slight
reduction of $\sigma_{\rm el}$ and $\kappa$ in Figure \ref{fig7}. However, 
the BGK collision term still retains the dominance over the RT 
collision term for both thermal and electrical conductivities, even
in the QPD of partons.

\begin{center}
\begin{figure}
\begin{tabular}{cc}
\includegraphics[width=6cm]{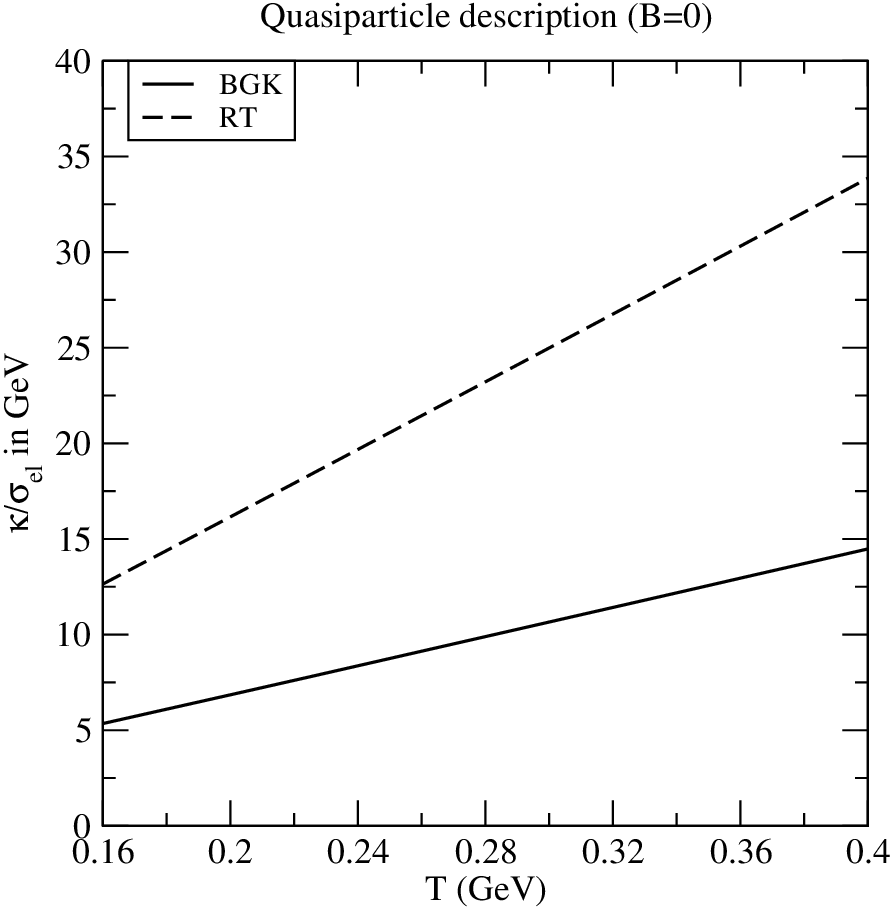}&
\hspace{1cm}
\includegraphics[width=6cm]{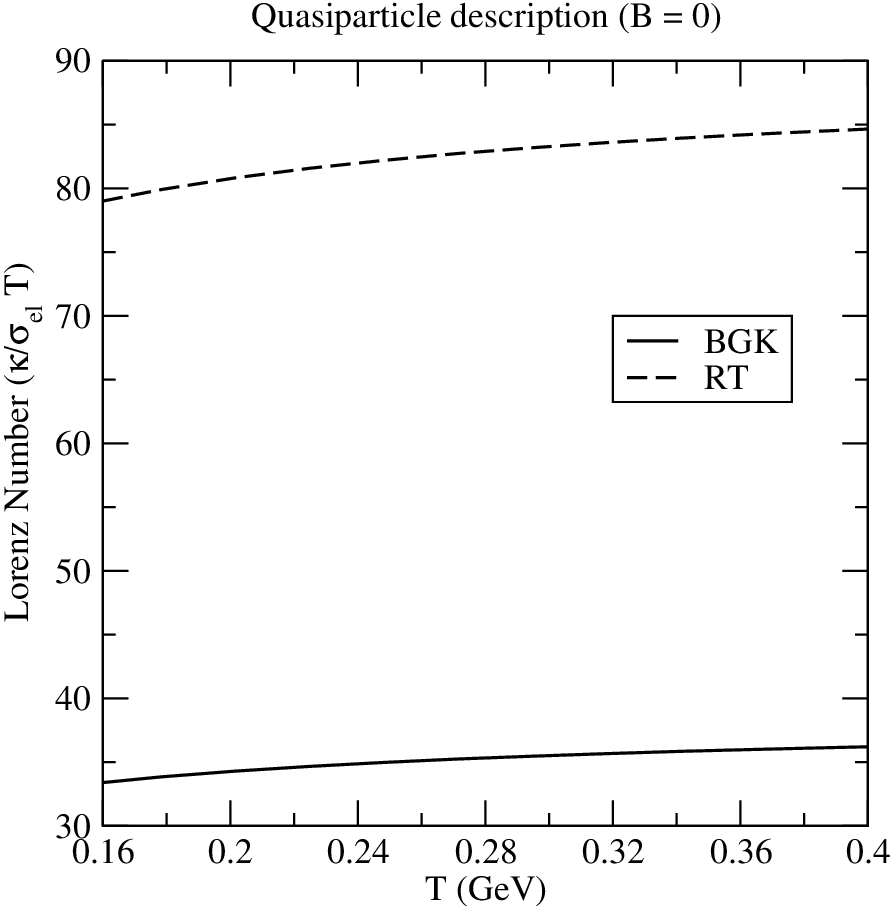}\\
(a)&(b)
\end{tabular}
\caption{Variation of the ratio of the thermal conductivity to the electrical conductivity  ($\kappa/\sigma_{el}$) and Lorenz number
with temperature.} 
\label{fig8} 
\end{figure}
\end{center}

Now we will see how the QPD of partons
affect the ratio, $\kappa/\sigma_{\rm el}$ (and the Lorenz number
as well) and the Knudsen number in the absence of magnetic field.
The aforesaid discussion on the charge and heat transport helps to 
understand the slight decrease in 
the ratio ($\kappa/\sigma_{\rm el}$) and the Lorenz number 
in Figure \ref{fig8} with respect to the noninteracting scenario
(in Figure \ref{fig2}).

\begin{center}
\begin{figure}[H]
\begin{tabular}{cc}
\includegraphics[width=6cm]{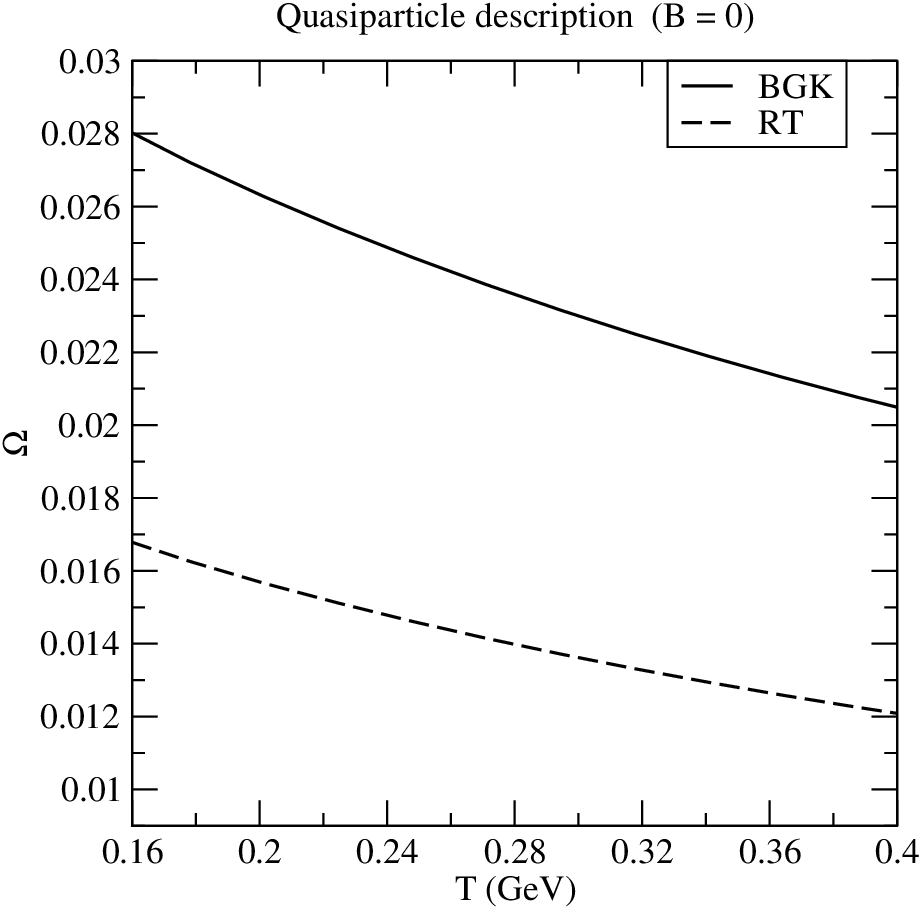}&
\hspace{1cm}
\includegraphics[width=6cm]{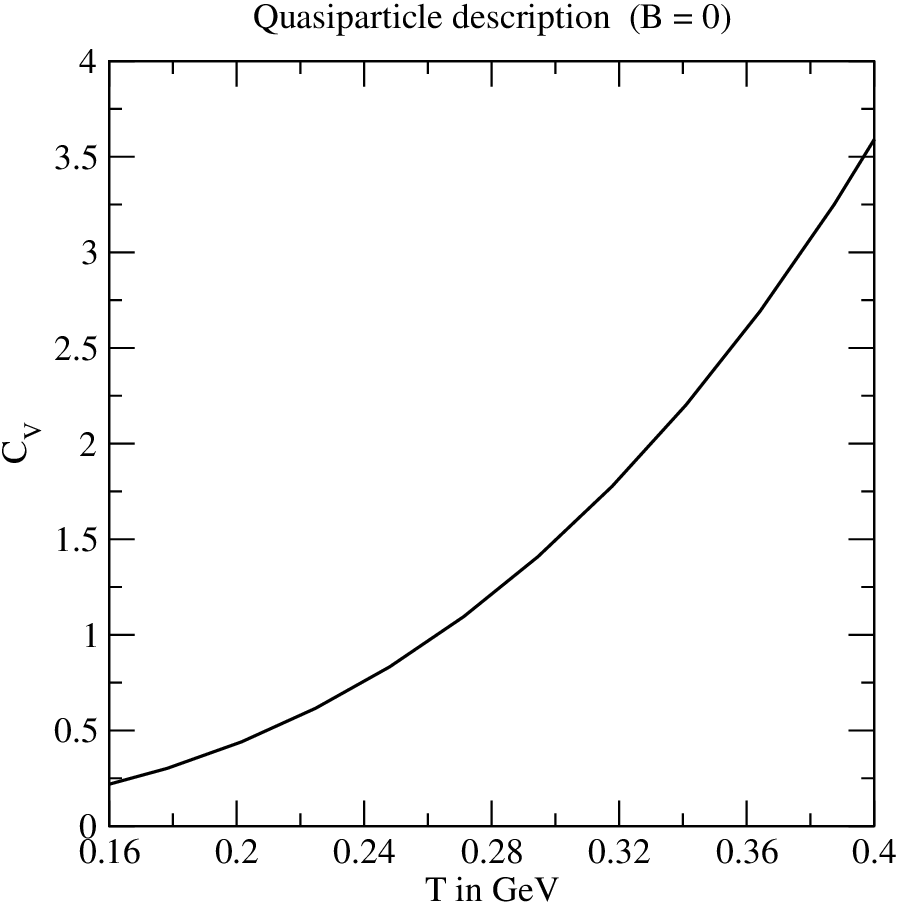}\\
(a)&(b)
\end{tabular}
\caption{Variation of the Knudsen number ($\Omega$) and specific heat
with temperature.} 
\label{fig9} 
\end{figure}
\end{center}

Similarly, the QPD of partons in the absence of magnetic field 
also reduces the Knudsen number a little bit, in comparison to 
noninteracting case (as in Figure \ref{fig3}(a)), 
which is mainly due to the opposite behaviour in thermal conductivity and 
specific heat in quasiparticle description (as seen in Figures
\ref{fig7}(b) and \ref{fig9}(b), respectively). The reduction
in $\Omega$ is in line with the fact that the more the 
interactions among the
constituents the quicker the system comes to local equilibrium. 

\subsection{Electrical and Thermal Conductivity in $B \neq 0$: Wiedemann Franz 
Law and Knudsen Number}
\begin{center}
\begin{figure}[H]
\begin{tabular} {c c}
\includegraphics[width=6cm]{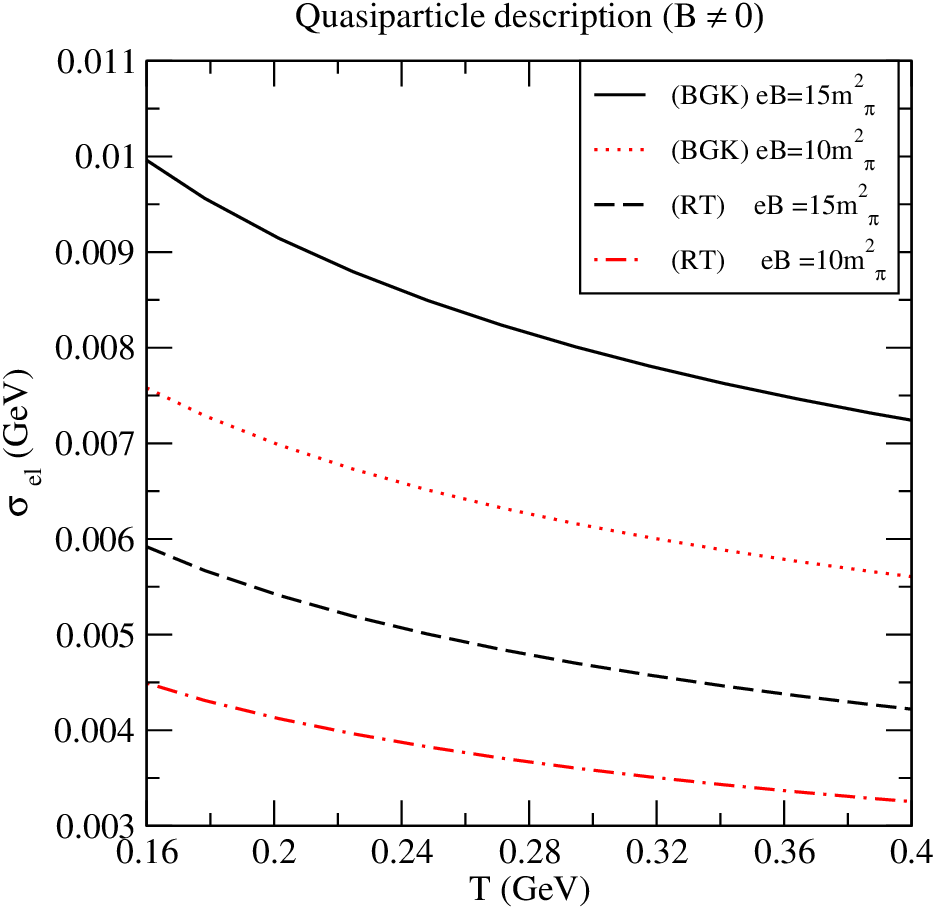}&
\hspace{1cm}
\includegraphics[width=6cm]{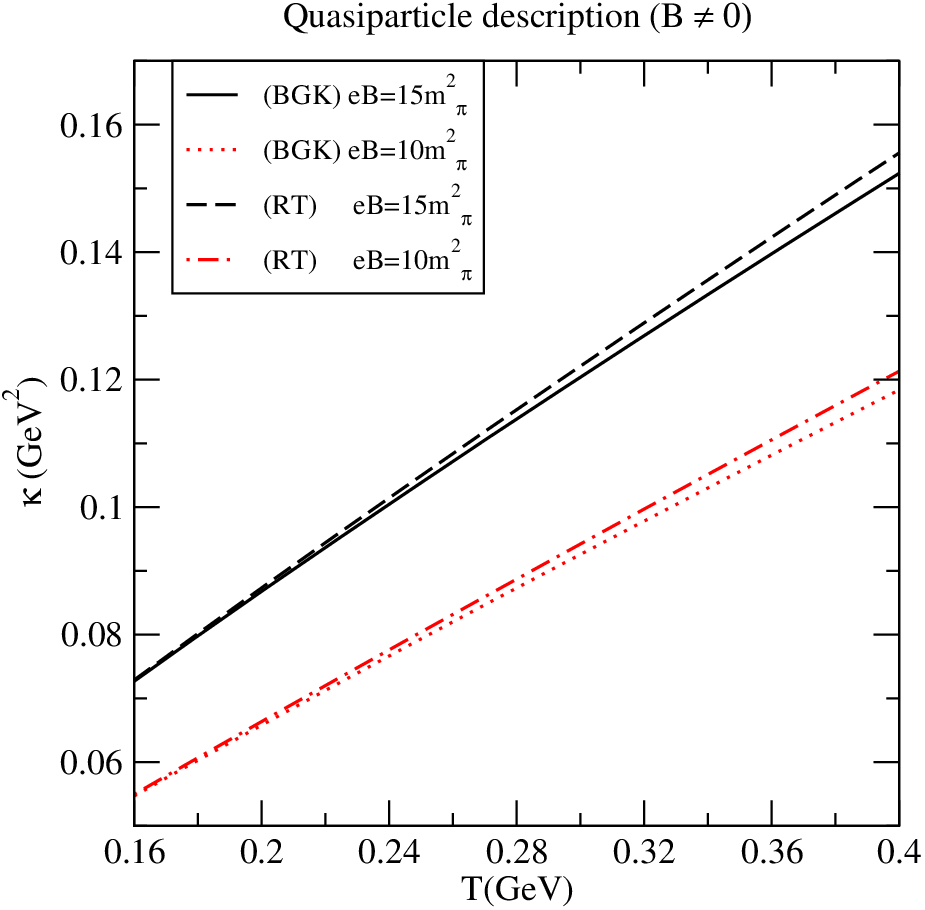}\\
(a) & (b)
\end{tabular} 
\caption {Variation of  electrical ($\sigma_{el}$) (left) and  
thermal ($\kappa$) (right) conductivities with the temperature.} 
\label{fig10}
\end{figure}
\end{center}

To see how the unusually larger values of $\sigma_{\rm el}$
and $\kappa$ for noninteracting partons in a strong $B$ (in 
Figure \ref{fig4}) could be affected by the QPD of partons,
we have now computed them in Figure \ref{fig10} with the in-medium
quark masses at finite temperature and strong $B$ from \eqref{massTB}.

We find that both $\sigma_{\rm el}$ and $\kappa$ gets
reduced by an order three, compared to the noninteracting  picture 
(Figure \ref{fig4}).
The primary reason behind this observation lies in the dispersion relation 
of partons in strong $B$, where the collective oscillation sets in 
much bigger scale than in the absence of $B$.
As a secondary reason, the dispersion relation, in turn, tames the 
relaxation-time, $\tau^B$ in strong $B$ (seen in Figure \ref{fig13}, which 
can be understood by the fact that the infra-red singularity in gluon-gluon 
cross-section 
in $\tau^B$ is cured by the the mass generated in the medium, which, 
in the presence of strong $B$, becomes much larger than in $B=0$ (seen 
in the Figure \ref{fig14}). 
\begin{table}[h]
\begin{center} 
\caption{Masses generated for $u$, $d$ and $s$ flavours in a thermal QCD medium
in the absence and presence of strong $B$ background.}
\begin{tabular}{|c|c|c|c|c|c|c|c|c|c|c|c|c|c|c|c|c|}
\hline 
&\multicolumn{3}{c|}{$m_i$ (B=0) } & 
\multicolumn{3}{c|}{$m_i$ (B $ \neq 0$)}\\
\cline{2-4}
\cline{5-7} 
\hline 
 &  u quark  & d quark  & s quark & u quark & d quark & s quark  \\ 
 & (GeV) & (GeV) & (GeV) &  (GeV) & (GeV) & (GeV) \\ 
\hline
 Low T & 0.1464  & 0.1485 &0.2275 & 0.7622 & 0.4187 & 0.1652 \\ 
 (0.16 GeV) & & & & & & \\
\hline
High T & 0.2935 & 0.2992 & 0.3735 & 1.2156 & 0.6669 & 0.2224  \\
(0.40 GeV) & & &  &&&\\
\hline 
\end{tabular}
\end{center} 
\end{table}

\begin{center}
\begin{figure}[H]
\begin{tabular}{cc}
\includegraphics[width=6cm]{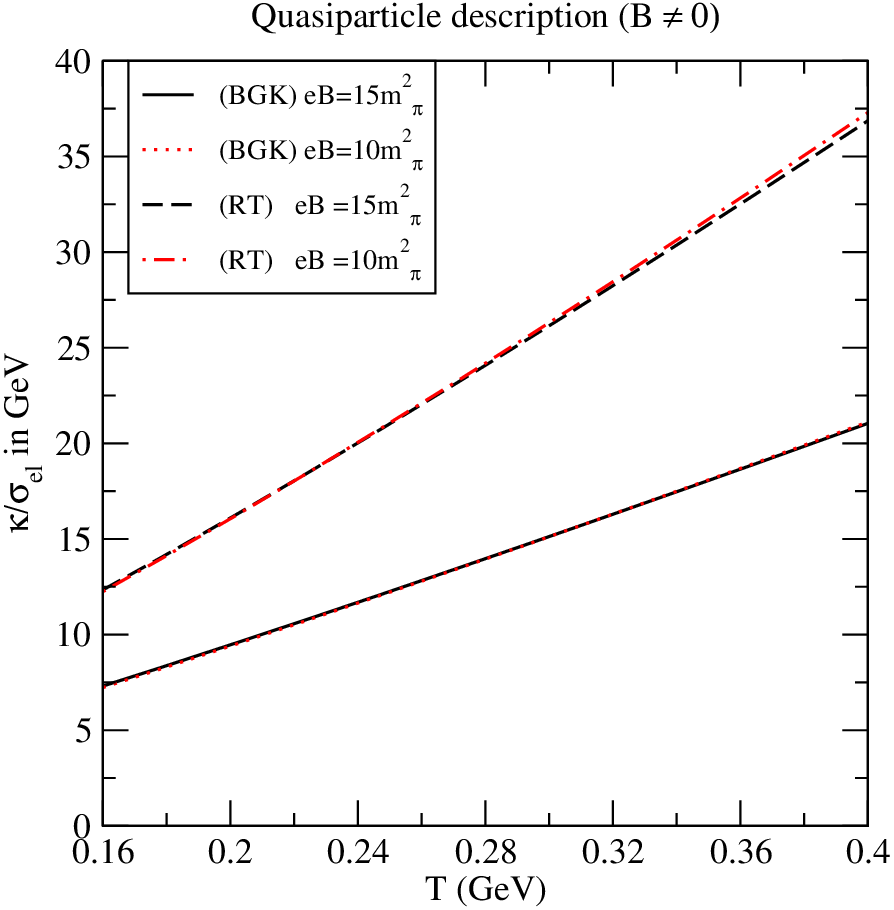}&
\hspace{1cm}
\includegraphics[width=6cm]{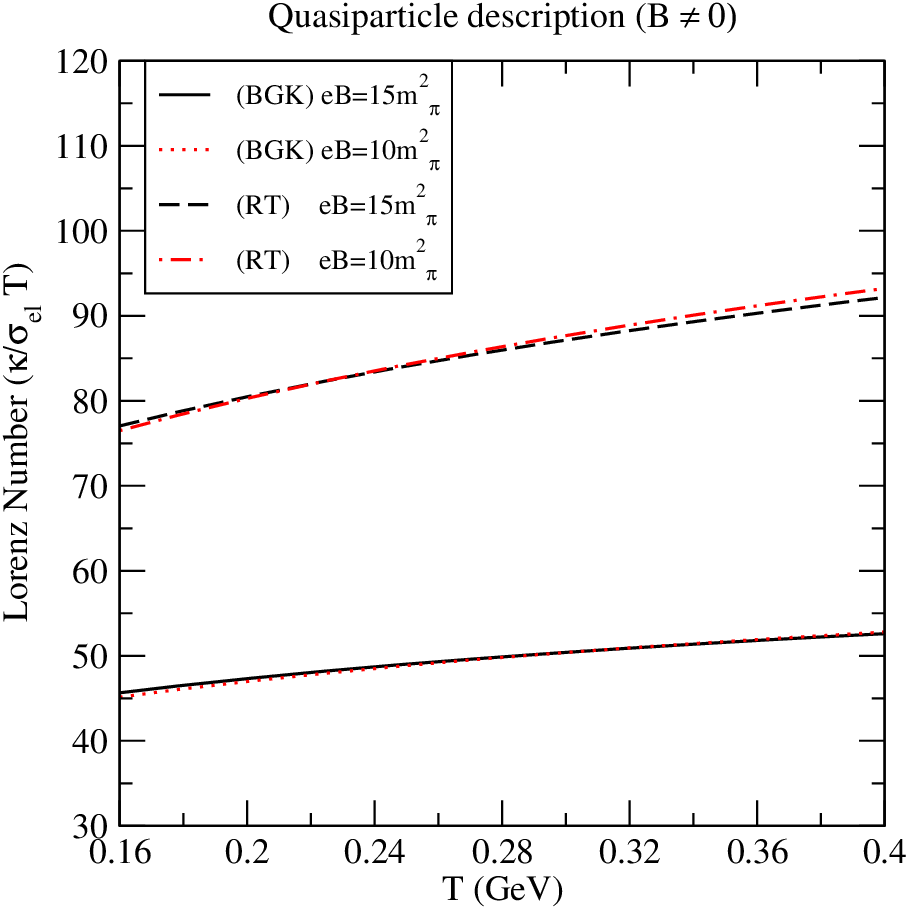}\\
(a)&(b)
\end{tabular}
\caption{Variation of the ratio of the thermal conductivity to the electrical conductivity  ($\kappa/\sigma_{el}$) and Lorentz number 
with temperature.} 
\label{fig11} 
\end{figure}
\end{center}

\begin{center}
\begin{figure}[H]
\begin{tabular}{cc}
\includegraphics[width=6cm]{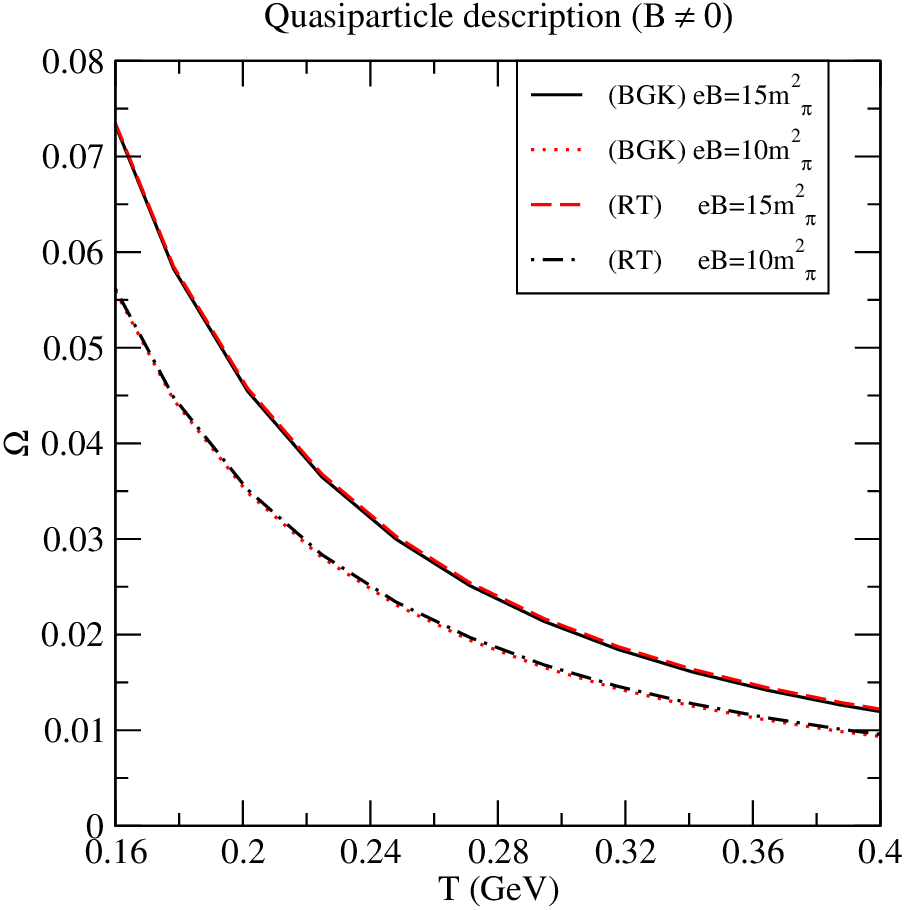}&
\hspace{1cm}
\includegraphics[width=6cm]{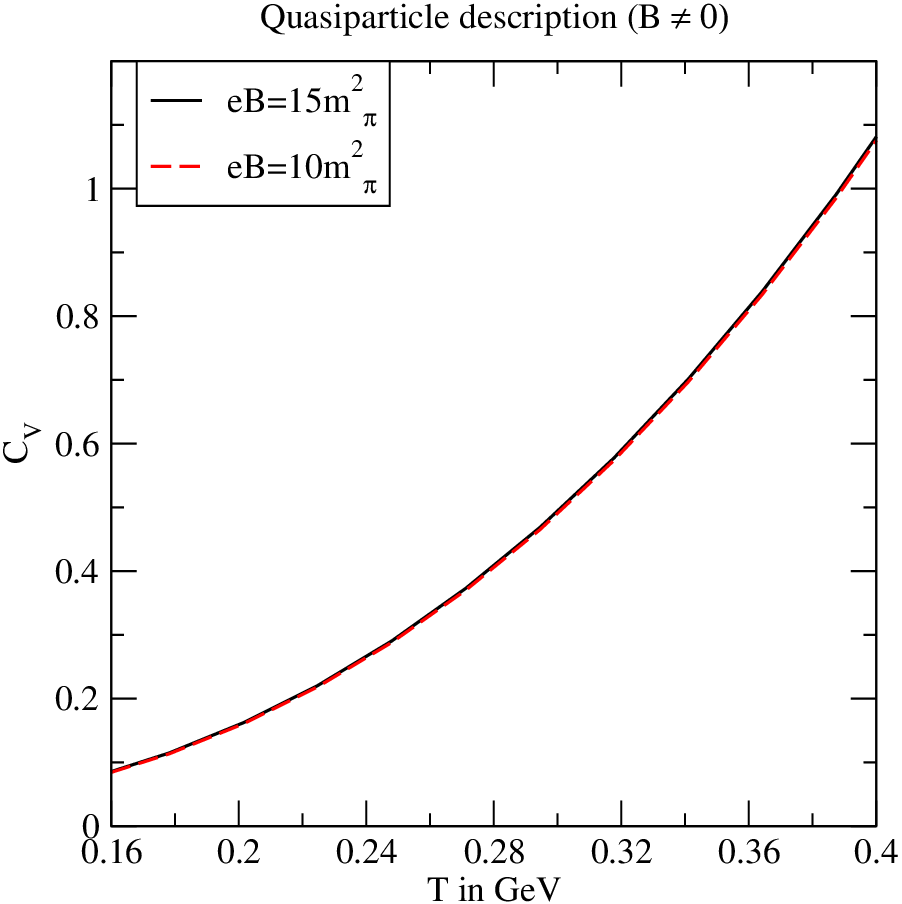}\\
(a)&(b)
\end{tabular}
\caption{Variation of the Knudsen number ($\Omega$) and specific heat
with temperature.} 
\label{fig12} 
\end{figure}
\end{center}

\begin{center}
\begin{figure}[H]
\begin{tabular}{cc}
\includegraphics[width=6cm]{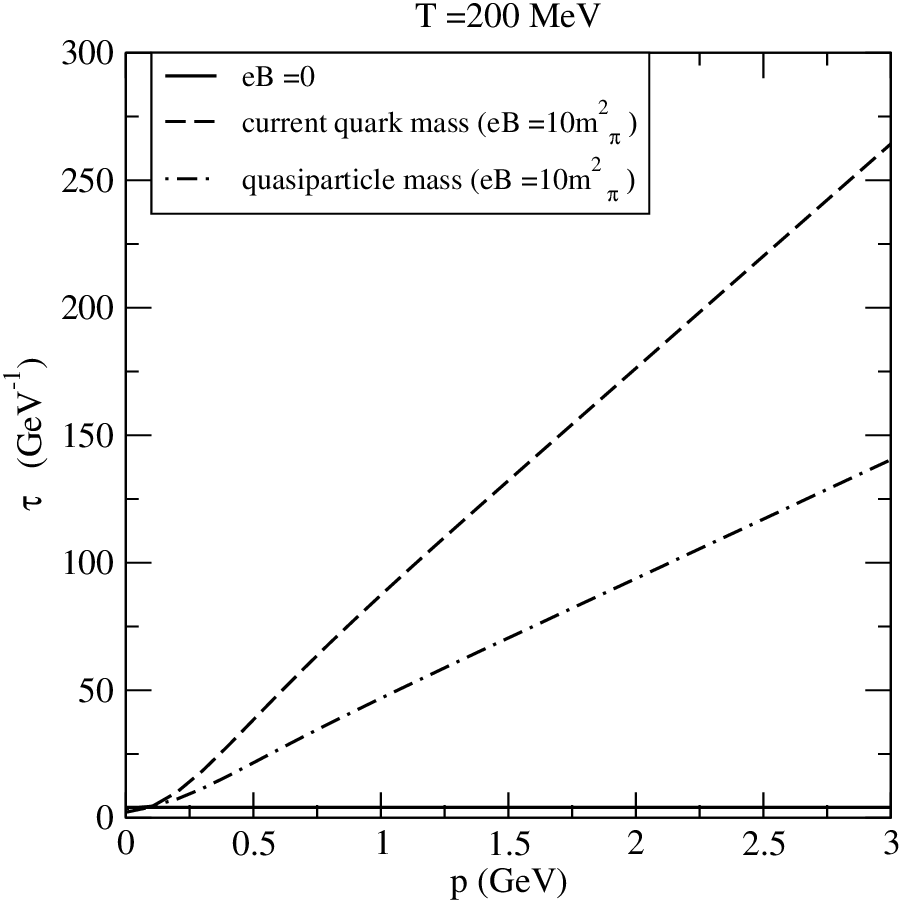}&
\hspace{1cm}
\includegraphics[width=6cm]{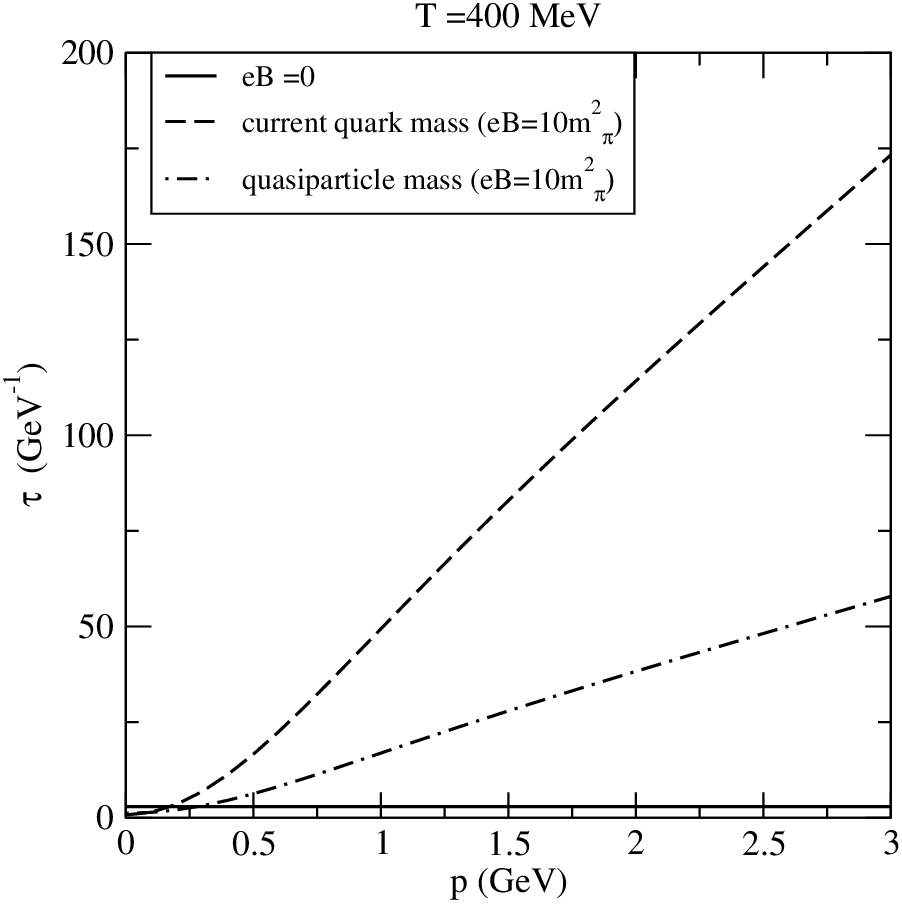}\\
(a)&(b)
\end{tabular}
\caption{Variation of the relaxation time (for s quark)
with momentum } 
\label{fig13}
\end{figure}
\end{center}

\begin{center}
\begin{figure}[H]
\begin{tabular}{cc}
\includegraphics[width=5cm]{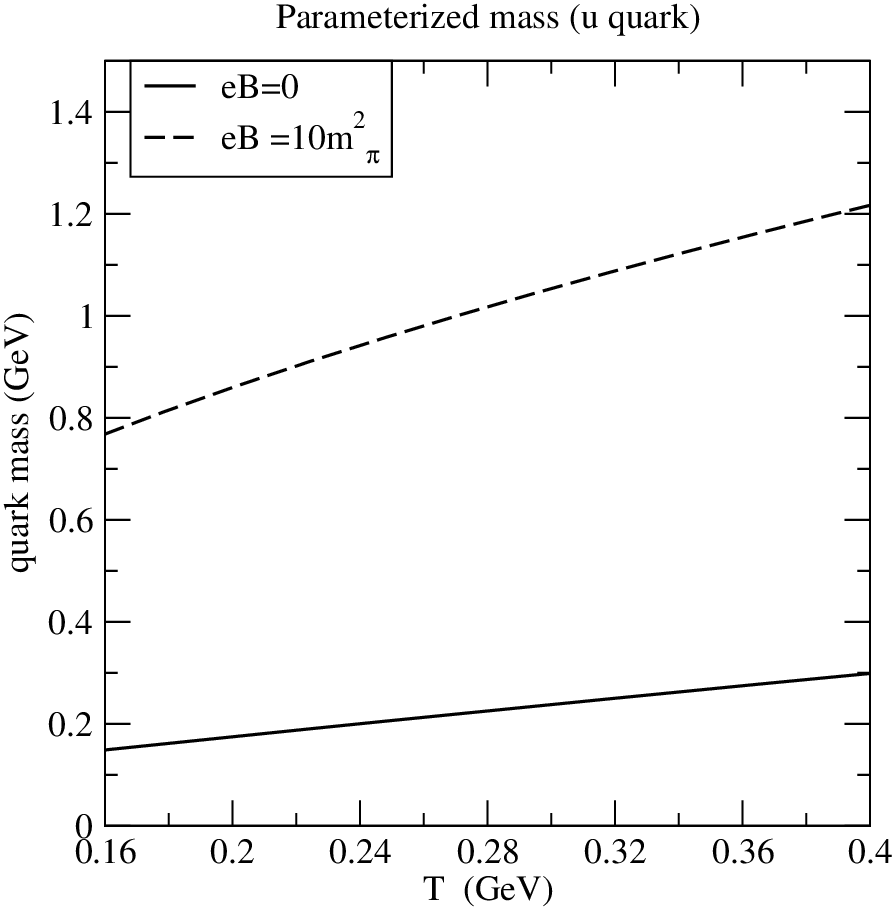}
\hspace{0.5cm}
\includegraphics[width=5cm]{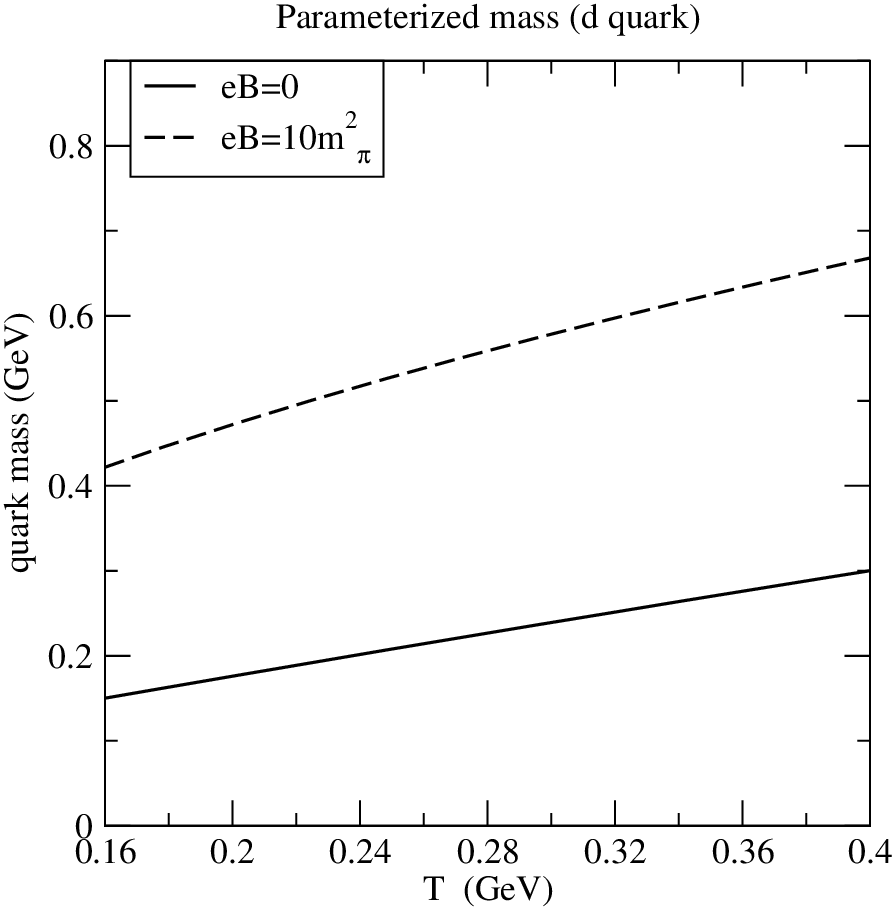}
\hspace{0.5cm}
\includegraphics[width=5cm]{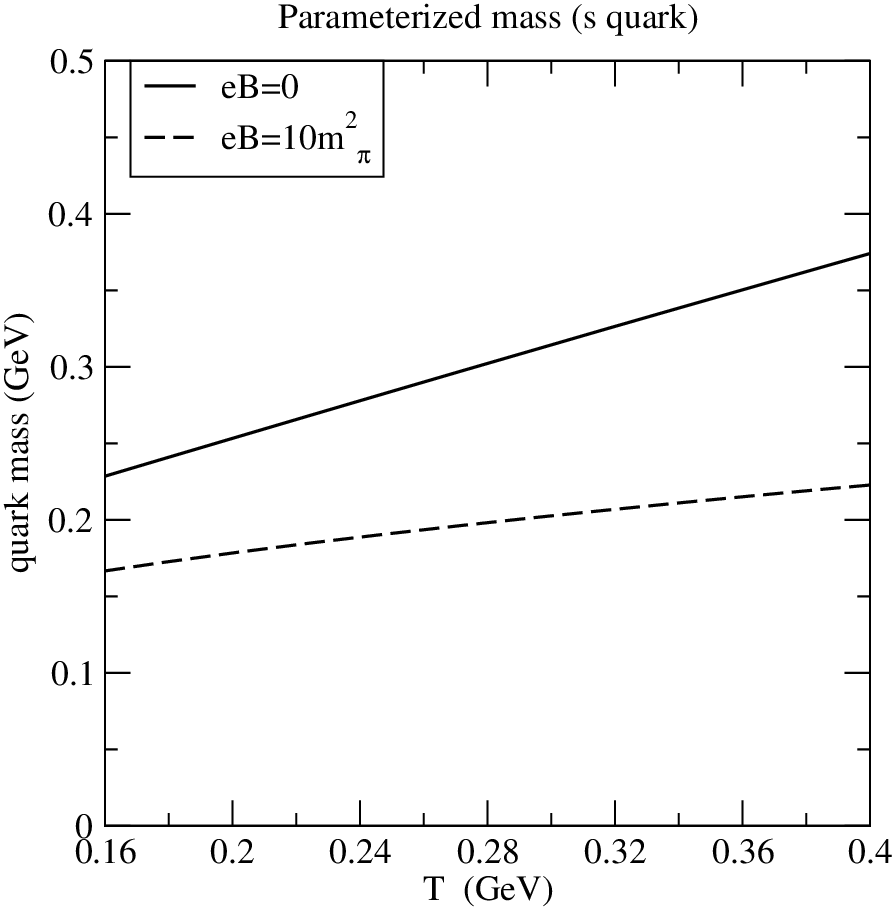}
\end{tabular}
\caption{Variation of parameterized quark masses with temperature}
\label{fig14} 
\end{figure}
\end{center}

Finally, the aforesaid behaviour of $\sigma_{\rm el}$ and $\kappa$
in strong $B$ facilitates to understand the effect of QPD
on the ratio, $\kappa/\sigma_{\rm el}$ and subsequently on the Lorenz number, 
$L_R$ in Figure \ref{fig11}(a) and Figure \ref{fig11}(b), respectively, wherein
both of them get amplified. Last but not the least, the best reward of
quasiparticle description gets reflected in the drastic
reduction of Knudsen number, $\Omega$, which looks now feasible (as seen 
in Figure \ref{fig12}(a)). Earlier in noninteracting description,
$\Omega$ in strong $B$ becomes much larger than one, implying that
the system runs away from  equilibrium. To understand
the abovementioned behaviour of $\Omega$, we also compute the  another
factor in $\Omega$, the specific heat
as a function of $T$ in the same description in Figure \ref{fig12}(b).
Finally, the excerpts of our exploration is that the QPD of partons almost 
smears the effect of the collision terms, {\em at least}, on the heat 
transport ($\kappa$) in strong $B$,
(as seen in \ref{fig10} (b)), which is being translated into a
interesting collision term dependence in the Lorenz number and
Knudsen number in Figure \ref{fig11} (b) and Figure \ref{fig12} (a),
respectively.

%

Finally, we have shown in Table 2 how the quasiparticle description affects 
the overall effect of collision term and the subsequent modulation of 
strong $B$ on the charge and heat transport.
\begin{table}[h]
\begin{center} 
\caption{Effect of collision term on the charge and heat transport 
in the absence and presence of strong magnetic field with the masses 
of $u$, $d$ and $s$ flavours generated in the thermal medium in
background of strong $B$.}
\begin{tabular}{|c|c|c|c|c|c|c|c|c|c|c|c|c|c|c|c|c|}
\hline 
&\multicolumn{6}{c|}{B=0}  &\multicolumn{6}{c|}{B $\neq$ 0}\\
\cline{2-7} 
\cline{8-13}
\hline 
Temperature &  $\frac{\sigma^{BGK}}{\sigma^{RT}}$ & 
$\frac{\kappa^{BGK}}{\kappa^{RT}}$ & 
$\frac{\frac{\kappa}{\sigma}^{BGK}}{\frac{\kappa}{\sigma}^{RT}}$ & 
$\frac{L_R^{BGK}}{L_R^{RT}}$ & 
$\frac{\Omega^{BGK}}{\Omega^{RT}}$ & 
$c_{{}_V}$ & 
$\frac{\sigma^{BGK}}{\sigma^{RT}}$ & 
$\frac{\kappa^{BGK}}{\kappa^{RT}}$ & 
$\frac{\frac{\kappa}{\sigma}^{BGK}}{\frac{\kappa}{\sigma}^{RT}}$ & 
$\frac{L_R^{BGK}}{L_R^{RT}}$ & 
$\frac{\Omega^{BGK}}{\Omega^{RT}}$ & 
$c_V$ \\ 
\hline
 Low temp. & 4.06  & 1.76 & 0.43 & 0.42 & 1.68 &0.20
  & 1.68 & $\approx$ 1.0 & 0.58 & 0.59 & $\approx$ 1.0 & 0.08  \\ 
 (0.160 GeV) & & & & & & & & & & & &\\
\hline
High temp. & 4.03  & 1.70 & 0.42 & 0.43 & 1.70  & 3.58 
& 1.71 & 0.98 & 0.57 & 0.57 & $\approx$1.0 & 1.07 \\
(0.400 GeV) & & & & & & & & & & & &\\
\hline 
\end{tabular}
\end{center} 
\end{table}

\section{Conclusion and future outlook}
Our aim is to study the charge and heat transport in a strongly interacting 
thermal QCD medium in a background of strong homogeneous magnetic field, 
through the respective transport coefficients. This 
is a complicated proposition to start with, so we have
adopted a bottom-to-top approach to handle the problem in a
kinetic theory approach. {\em First of all}, we have checked how the modified
BGK collision integral, unlike the commonly employed
collision terms of relaxation type (RT), affects the charge 
and heat transport in a thermal QCD medium.
This exploration is germinated due to the fact that the BGK collision term 
ensures conservation of 
particle number, momentum and energy in each collision, {\em unlike} 
the usually adopted
collision terms of relaxation type, where the conservation of 
particle is ensured only on the average of a cycle.
{\em Secondly}, we see that how an ambient strong magnetic field (may
be produced in the peripheral events of ultrarelativistic heavy-ion 
collisions) modulates the collision integral, which, in turn, affects
the aforesaid heat and charge transport. This exploration is motivated
by the fact that the strong $B$ strongly affects the phase space, 
and the collision time too, which will ultimately affect the 
solution of the Boltzmann equation. Thirdly, instead
of independent particle excitations, we see how the collective oscillation of 
the medium through the quasiparticle description of partons
affects the occupation probability, which, in turn, affects the charge 
and heat transport coefficients and their derived coefficients. 

The modified BGK collision integral enhances both charge and heat transport,
{\em especially} more to the charge transport, compared to RT collision 
integral. As a consequence, the RT collision integral dominates the 
ratio of thermal-to-electrical conductivity (and the 
Lorenz number, $L_R$) whereas the BGK collision integral 
dominates the equilibration through the Knudsen 
number ($\Omega$), for  B=0. However, 
in the presence of strong $B$, 
both electrical and thermal conductivities get amplified but 
the collision integrals affect on charge and heat transport differently.
To be specific, BGK collision integral still dominates the charge 
transport whereas RT collision integral dominates the heat transport and
overall strong $B$ smears the effect of collision integral.
 However, the large values of thermal
conductivity and the reduction of the specific heat
due to the dimension reduction make the equilibration factor, Knudsen 
number unusually large, which defies physical interpretation.
Finally the quasiparticle description of the partons in the absence
of strong $B$ impedes both charge and heat transport, which
is reflected in the slight decrease of the conductivities. However,
the quasiparticle description in strong $B$
makes the transport phenomena interesting (as seen in Table 2): 
i) The large values of 
$\sigma_{el}$ and $\kappa$ are tamed to the physically plausible 
values, ii) The effect of collision integral is no more sensitive to
the heat transport, except $T$ is very large, and iii) the characteristic 
$T$-dependence of conductivities get reversed, {\em namely} 
$\sigma_{\rm el}$ now 
decreases with $T$ and the increase of $\kappa$ with $T$ is linear.
As a consequence, $L_R$ gets bigger 
and increases rapidly with $T$ and becomes almost independent of $B$.


In future we are planning to explore the above study to the momentum
transport and the affiliated coefficients associated to the
momentum transport and investigate its implications 
in heavy-ion phenomenology by studying the 
hydrodynamic evolution of the medium along with strong
$B$ produced in ultrarelativistic heavy ion collisions.

\section*{Acknowledgements}
One of us (B. K. P.) is thankful to the Council of Scientific and 
Industrial Research (Grant No. 03(1407)/17/EMR-II) for the financial assistance.

\appendix
\appendixpage
\addappheadtotoc
\begin{appendix}
\renewcommand{\theequation}{A.\arabic{equation}}
\section{Derivation of equation \eqref{appendix_A}}
Putting the partial derivatives from  \eqref{partial11} and 
\eqref{partial22}  in \eqref{rbte_sigmatemp}, we get
\ba 
-q_i \beta {\bf E.p} f_{\mathrm eq,i}
(1-f_{\mathrm eq,i})-
q_i \beta p_0 \frac{{\bf E. p}}{\omega_i}
 f_{\mathrm eq,i}(1-f_{\mathrm eq,i})&=&C[f],
\label{aa1} 
\ea
where the  BGK collision term 
is given by  \eqref{BGK_thermal} as
\ba
C[f]&=&-p^{\mu}u_{\mu}\nu_i \left(f_i-n_{{}_i}n_{\mathrm eq,i}^{-1} f_{\mathrm eq,i}
\right)\nonumber\\
&=&-p^{\mu}u_{\mu}\nu_i\left(f_i-\frac{g_i \int_p (f_{\mathrm eq,i}+\delta f_i)}{n_{\mathrm eq,i}}
f_{\mathrm eq,i}\right)\nonumber\\
&=&-p^{\mu}u_{\mu}\nu_i\left(f_i-\frac{(g_i \int_p f_{\mathrm eq,i}
+g_i\int_p\delta f_i)}{n_{\mathrm eq,i}}f_{\mathrm eq,i}\right)\nonumber\\
&=&-p^{\mu}u_{\mu}\nu_i\left(\delta f_i-g_i n_{\mathrm eq,i}^{-1}
f_{\mathrm eq,i}\int_p\delta f_i\right).
\ea 
Now putting the value of $C[f]$ in \eqref{aa1} we get
\ba
2q_i\beta \tau_i~ \frac{{\bf E.p}}{\omega_i} f_{\rm eq,i}\left( 
1-f_{\rm eq,i} \right)&=&\left(\delta f_i
-g_i n_{\mathrm eq,i}^{-1}f_{\mathrm eq,i}\int_p\delta f_i\right)
\ea

\renewcommand{\theequation}{B.\arabic{equation}}
\section{Derivation of the infinitesimal deviation ($\delta f_i^B$) from 
the local equilibrium}
In order to find the deviation ($\delta f_i^B$) in the 
distribution function, we solve relativistic Boltzmann equation 
\eqref{RBTE_heat}. 
Substituting the partial derivatives from  
\eqref{partial2} and 
\eqref{partial3} on the left hand side (denoted as LHS)
 of eqn. \eqref{RBTE_heat},
 we get
\begin{eqnarray}
{\rm LHS}&=&(p^{0}\partial_{0}T+p^{3}\partial_{3}T)\frac{\partial f_i^{B}}{\partial T}
+(p^{\mu} u_{\nu}\partial_{\mu}p^{\nu}+p^{\mu} p^{\nu}\partial_{\mu}u_{\nu})
\frac{\partial f_i^{B}}{\partial p^0}+q_i\left[F^{03}p_3
\frac{\partial f_i^{B}}{\partial p^0}+F^{30}p_0\frac{
\partial f_i^{B}}{\partial p^3}
\right],\nonumber \\
&=&\frac{p^0}{T}f_{\rm eq,i}^{B}(1-f_{\rm eq,i}^{B})\left[\frac{1}{T}(p^{0}
\partial_{0}T+p^{3}\partial_{3}T)
-\frac{1}{p^0}(p^{\mu} u_{\nu}\partial_{\mu}p^{\nu}
+p^{\mu} p^{\nu}\partial_{\mu}u_{\nu})-2q_i
\frac{E_3p_3}{p^0}\right],\nonumber\\
&=&\frac{p^0}{T}f_{\rm eq,i}^{B}(1-f_{\rm eq,i}^{B})\left[\frac{1}{T}(p^{0}
\partial_{0}T+p^{3}\partial_{3}T)
-\frac{1}{p^0}(p^{0} \partial_{0}p^{0}+p^{3}
 \partial_{3}p^{0})
 -\frac{1}{p^0}(p^{0} p^{\nu}\partial_{0}u_{\nu}+p^{3} p^{\nu}
\partial_{3}u_{\nu})\right.\nonumber\\
&&\left. -2q_i\frac{E_3p_3}{p^0}\right]\nonumber\\
&=&\frac{p^0}{T}f_{\rm eq,i}^{B}(1-f_{\rm eq,i}^{B})\left[\frac{1}{T}
(p^{0}\partial_{0}T+p^{3}\partial_{3}T)
+T\partial_0\left(\frac{\mu}{T}\right)+\frac{T}
{p^0}p^3\partial_3\left(\frac{\mu}{T}\right)
 -\frac{1}{p^0}(p^{0} p^{\nu}\partial_{0}
u_{\nu}+p^{3} p^{\nu}\partial_{3}u_{\nu})\right.\nonumber\\
&&\left. -\frac{2q_iE_3p_3}{p^0}\right]
\end{eqnarray}
substituting $\partial_3\left(\frac{\mu_i}{T}\right)=-\frac{h_i^B}{T^2}
\left(\partial_3T-\frac{T}{n^B_{eq,i}h_i^B}\partial_3P\right)$ and $\partial_0u_{\nu}=\frac{\nabla_{\nu}P}{n^B_{eq,i}h_i^B}$ from the energy momentum conservation
on the left hand side (LHS) and BGK collision term on the right hand side
of \eqref{RBTE_heat}
\begin{eqnarray}
\frac{p^0}{T}f_{\rm eq,i}^{B}(1-f_{\rm eq,i}^{B})\left[\frac{p^0}{T}\partial_{0}T
+\frac{(p^0-h_i^B)}
{p^0}\frac{p^3}{T}\left(\partial_3T-\frac{T}{n^B_{eq,i}h_i^B}\partial_3P\right)
+T\partial_0\left(\frac{\mu}{T}\right)
-\frac{p^3p^{\nu}}{p^0} \partial_{3}u_{\nu}\right.\nonumber\\
\left. -2q_i\frac{E_3p_3}{p^0}\right]
=-p^{\mu}u_{\mu}\nu_i^ B\left(\delta f_i^B-
g_i{n_{\mathrm eq,i}^{B}}^{-1}f_{\rm eq,i}^B\int \delta f_i^B\right),
\label{rbte_heat}
\end{eqnarray}
which can further be solved for $\delta f_i^B$ upto first order as
\begin{eqnarray}
\delta f_i^{B}=\delta f_i^{B(1)}+g_i{n_{\mathrm eq,i}^{B}}^{-1}f_{\rm eq,i}^B
\int_{p'} \delta f_i^{B(1)},
\end{eqnarray}
where 
\begin{eqnarray}
\delta f_i^{B(1)}&=&-\frac{f_{\rm eq,i}^{B}(1-f_{\rm eq,i}^{B})\tau_i^{B}}{T}
\left[\frac{p^0}{T}\partial_{0}T+\frac{(p^0-h_i^B)}
{p^0}\frac{p^3}{T}\left(\partial_3 T-\frac{T}{n_{eq,i}^Bh_i^B}
\partial_3P\right)
+T\partial_0\left(\frac{\mu}{T}\right)\right.\nonumber \\
&&\left. -\frac{p^3p^{\nu}}{p^0} \partial_{3}u_{\nu}
-2q_i\frac{E_3p_3}{p^0}\right].
\end{eqnarray}

\renewcommand{\theequation}{B.\arabic{equation}}
\section{Calculation of the quark self energy}
The transverse component of the momentum in the quark propagator
becomes very small $(k_{\perp}\approx 0)$, so the exponential factor
($e^{-k_{\perp}/|q_iB|}$) in \eqref{quark_prop}  becomes unity.
The quark self energy \eqref{quark_self} in the strong magnetic 
field can be written as
\begin{eqnarray}
\Sigma(p_\parallel) &=& \frac{2g^2}{3\pi^2}|q_iB|T\sum_n\int dk_z\frac{\left[\left(1+\gamma^0\gamma^3\gamma^5\right)\left(\gamma^0k_0
-\gamma^3k_z\right)-2m_i\right]}{\left[k_0^2-\omega^2_k\right]\left[(p_0-k_0)^2-\omega_{pk}^2\right]}\nonumber \\ 
&=& \frac{2g^2|q_iB|}{3\pi^2}\int dk_z\left[(\gamma^0+\gamma^3\gamma^5)L^1-(\gamma^3+\gamma^0\gamma^5)k_zL^2\right]
\label{quark_self1}
,\end{eqnarray}
where $\omega_k^2=k_z^2+m_i^2$, $\omega_{pk}^2=(p_z-k_z)^2$
and $L^1$ and $L^2$ are the frequency sum which are given by
\begin{eqnarray}
&&L^1=T\sum_nk_0~\frac{1}{\left[k_0^2-\omega_k^2\right]}\frac{1}{\left[(p_0-k_0)^2-\omega_{pk}^2\right]} ~, \\ 
&&L^2=T\sum_n\frac{1}{\left[k_0^2-\omega_k^2\right]}\frac{1}{\left[(p_0-k_0)^2-\omega_{pk}^2\right]}
\end{eqnarray}
after summing the above frequency sum the self energy \eqref{quark_self1}
becomes
\begin{eqnarray}
\Sigma(p_\parallel)&=&\frac{g^2|q_iB|}{3\pi^2}\int \frac{dk_z}{\omega_k}
\left[\frac{1}{e^{\beta\omega_k}-1}+\frac{1}{e^{\beta\omega_k}+1}\right]\nonumber\\
&&\times\left[\frac{\gamma^0p_0}{p_\parallel^2}+\frac{\gamma^3p_z}{p_\parallel^2}+\frac{\gamma^0\gamma^5p_z}{p_\parallel^2}+\frac{\gamma^3\gamma^5p_0}{p_\parallel^2}\right],
\end{eqnarray}
which after integration over $k_z$ simplified into
\begin{eqnarray}
\Sigma(p_\parallel)=\frac{g^2|q_iB|}{3\pi^2}\left[\frac{\pi T}{2m_i}-\ln(2)\right]\left[\frac{\gamma^0p_0}{p_\parallel^2}+\frac{\gamma^3p_z}{p_\parallel^2}+\frac{\gamma^0\gamma^5p_z}{p_\parallel^2}+\frac{\gamma^3\gamma^5p_0}{p_\parallel^2}\right]
\label{quark_self2}
.\end{eqnarray}
\end{appendix}

\end{document}